\documentclass{ws-mplb}
\usepackage[super,sort,compress]{cite} 
\usepackage{xcolor}
\usepackage[verbose,hypertexnames=false]{hyperref}
\hypersetup{colorlinks=false,allbordercolors=blue,pdfborderstyle={/S/U/W 1}}
\usepackage{algorithm}      
\usepackage{algpseudocode}  
\usepackage{graphicx}      
\usepackage{subcaption}

\usepackage{tabularx}  
\usepackage{booktabs} 
\usepackage{placeins}
\usepackage{float}

\usepackage{natbib}
\usepackage{subcaption}

\begin{document}

\markboth{X.M. Yang,  J.Y. Jia and Z.L. Deng}{A Bayesian approach with persistent homology prior for Robin coefficient identification in parabolic problem }

%
%

\title{A Bayesian approach with persistent homology prior for Robin coefficient identification in a parabolic problem 
}

\author{Xiaomei Yang\footnote{Corresponding author}}

\address{School of Mathematics, Southwest Jiaotong University, 
Chengdu, 611756, Sichuan, China
\\
yangxiaomath@swjtu.edu.cn}

\author{ Jiaying Jia}

\address{School of Mathematics, Southwest Jiaotong University, 
Chengdu, 611756, Sichuan, China\\
jiaying@my.swjtu.edu.cn}

\author{ Zhiliang Deng}

\address{School of Mathematical Sciences, University of Electronic Science andTechnology of China, Xiyuan Ave, Chengdu, 611731, Sichuan, China\\
dengzhl@uestc.edu.cn.}

\maketitle


\begin{abstract}

The reconstruction of time-dependent Robin coefficients is a challenging inverse heat transfer problem due to its inherent ill-posedness. This paper introduces a hierarchical Bayesian approach integrated with a persistent homology (PH) prior for robust coefficient estimation. By quantifying the birth and death of topological features, the PH-based prior provides a global structural constraint that transcends local derivative-based penalties. Numerical experiments show that this topological perspective allows for the preservation of complex temporal profiles without the typical staircase distortions of total variation (TV) priors or the excessive blurring of Gaussian models. A key feature of our framework is the hierarchical implementation, which yields an automated, data-driven selection of hyperparameters. The results demonstrate that while PH-based inference yields competitive accuracy compared to TV regularization, it offers superior performance in preserving the multiscale characteristics of the Robin coefficient, providing a robust alternative for convective heat transfer diagnostics.
\end{abstract}

\keywords{Bayesian inference; persistent homology; Robin coefficient; inverse problem; parabolic equation; hierarchical Bayesian.}


\section{Introduction}\label{s:1}

The Robin coefficient characterizes the intensity of heat exchange in convective heat transfer and reflects the physical behavior near boundaries \cite{Alessandrini2003, Arthern2010, Inglese1997, Jin2009}.
In many practical applications, however, this coefficient cannot be measured or accessed directly.
As a result, inferring the Robin coefficient from indirect observations and quantifying the associated uncertainty become important and challenging tasks.
In this paper, we consider the following parabolic problem with an unknown Robin coefficient $\gamma(t)$: 
\begin{equation}\label{eqn1}
\begin{cases}
\frac{\partial u}{\partial t}(x, t)
- \frac{\partial}{\partial x}\!\left(\alpha(x)\frac{\partial u}{\partial x}(x, t)\right)
= f(x, t),
& x \in (0, 1), \ t>0, \\[1ex]
\alpha(0) \dfrac{\partial u}{\partial x}(0, t) = h_0(t),
& t>0, \\[2ex]
\alpha(1) \dfrac{\partial u}{\partial x}(1, t) + \gamma(t) u(1, t) = h_1(t),
&  t>0, \\[1ex]
u(x, 0) = g(x),
& x \in (0, 1).
\end{cases}
\end{equation}
Throughout this paper, all other coefficients, source terms, and boundary/initial data are assumed to be known.
Since the Robin coefficient $\gamma(t)$ cannot be accessed directly, its estimation relies on indirect observations of the state variable.

The inverse problem of identifying the Robin coefficient has been extensively studied in the literature.
On the theoretical side, various results concerning existence, uniqueness, and stability have been established; see, for example, \cite{Bellassoued2008, Biegert2009, Choulli1999, Jin2012, Shidfar2006, Slodicka2002, Slodicka2010}.
Beyond theoretical analysis, a wide range of numerical methods has been developed for the identification of Robin coefficients. Onyango et al.~\cite{Onyango2009} employed the boundary element method for coefficient inversion. Yang et al.~\cite{Yang2009} applied a conjugate gradient approach to estimate the Robin coefficient. Fang et al~\cite{Fang2009} combined boundary integral equation methods with regularization techniques to reconstruct spatially dependent coefficients. Jin et al \cite{Jin2012} considered space--time dependent Robin coefficients using gradient-based optimization under Tikhonov regularization. More recently, da Silva et al.~\cite{daSilva2021} integrated the method of fundamental solutions with a sequential importance resampling particle filter to estimate a time-dependent heat transfer coefficient.

In recent years, the Bayesian framework has gained increasing attention in the identification of Robin coefficients. Bayesian inference is well suited to quantify uncertainty and to handle complex data. Many prior models have been proposed. Among them, \cite{Jin2008, Jin2008withZou, Shi2023, Yan2009} use the Markov random field (MRF);  Rasmussen et al. \cite{Rasmussen2024} employ Matérn and squared-exponential Gaussian priors; and Yao et al.\cite{Yao2016} propose a TV–Gaussian hybrid prior that can capture sharp jumps in the target. In addition, \cite{Jin2008,  Shi2023, Yan2009} adopt a hierarchical Bayesian method with Gaussian priors to select the regularization parameter automatically.

Building on the existing studies, we adopt a prior constructed on the topological tool of persistent homology (called PH prior). This prior was originally proposed in \cite{Deng2025} for Bayesian inversion of potential coefficients. And to the best of our knowledge, it has not yet been applied to the identification of Robin coefficients.  For the PH-Gaussian hybrid prior \cite{Deng2025}, there is not a concrete strategy for selecting the regularization parameter $\lambda$.   In this paper, we address this issue by incorporating a hierarchical Bayesian framework to infer $\lambda$ from the data. By quantifying and encoding the topological features of the unknown function, the PH prior effectively preserves both global structures and local discontinuities during the inversion process. Compared with Gaussian priors, the PH-Gaussian hybrid prior shows better performance in handling target functions with sharp discontinuities or complex structures; at the same time, it preserves richer topological information than the TV prior.

The remainder of this paper is organized as follows. Section 2 describes fundamental ideas of Bayesian inference with an emphasis on PH prior and hierarchical Bayesian models. Section 3 provides numerical examples to validate the effectiveness of the proposed approach. Section 4 concludes the paper.

\section{The Bayesian approach with the PH-Gaussian prior }\label{s:2}


In this section, we present the Bayesian inversion for identifying the Robin coefficient using a PH–Gaussian prior. Before we introduce the Bayesian approach, we need know  persistent homology on 1D continuous function.

\subsection{Persistent homology on 1D continuous function}

In the inverse problem of identifying the one-dimensional Robin coefficient $\gamma(x)$, persistent homology (PH) is introduced to characterize the structural profile of the function from a topological perspective. By transforming the continuous functional graph into an algebraic topological object, this technique captures the birth and death of connected components through a filtration process.  This process yields a quantified topological characterization that effectively encodes the function's structural features.
This section introduces the persistent homology which forms a concrete basement for
our prior. For more details one can refer to \cite{Edelsbrunner2014, Kaczynski2004, Miller2021, Zheng2015, Zomorodian2005}.


We briefly review simplicial complexes and provide a formal definition for persistent homology.

\begin{definition}[k-simplex]

a \( k \)-simplex is the convex hull of  \( k+1 \) affinely independent points in some Euclidean space \( \mathbb{R}^n \) where \( n \ge k \).
\[
\Delta^k = \left\{(x_0, x_1, \dots, x_k) \in \mathbb{R}^{k+1} \ \middle| \ x_i \ge 0 \text{ for all } i, \text{ and } \sum_{i=0}^k x_i = 1\right\}
\]
\end{definition}
A simplicial subdivision of a curve is composed exclusively of 0-simplices and 1-simplices, representing its vertices and edges respectively.

\begin{definition}[Simplicial Complex]
A finite set of simplices $K$ in $\mathbb{R}^n$ is called a simplicial complex if it satisfies the following two conditions:
\begin{enumerate}
    \item {Face Enumeration:} Every face of any simplex $\sigma \in K$ is also an element of $K$.
    \item {Intersection Property:} For any two simplices $\sigma_1, \sigma_2 \in K$, their intersection $\sigma_1 \cap \sigma_2$ is either empty or a common face of both.
\end{enumerate}
\end{definition}

\begin{definition}[$p$-chain]
Let $K$ be a simplicial complex. A $p$-chain is a formal linear combination of all $p$-simplices in Simplicial Complex $K$, expressed as:$$c = \sum_{i} a_i \sigma_i, \quad a_i \in \mathbb{Z}$$where each $\sigma_i$ is a $p$-simplex in $K$.
\end{definition}

The set of all $p$-chains forms an Abelian group under addition defined component-wise,  denoted by $C_p(K)$.
Specifically, $C_0(K)$ and $C_1(K)$ are the free Abelian groups defined over the vertices $\{v_i\}$ and the directed edges $\{e_j\}$, respectively. These groups are connected by the boundary operator
$$\partial_p: C_p(K) \to C_{p-1}(K).$$
When $p=1$, the operator $\partial_1$ maps each oriented edge to the formal difference of its endpoints.
We also know that $\text{im}(\partial_{p+1}) $ is a subgroup of $\ker(\partial_p)$.
\begin{definition}[Homology group]
The $p$-th homology group of the simplicial complex $K$ is then defined as the quotient group
$$
H_p(K)=\operatorname{Ker}\left(\partial_p\right) / \operatorname{Im}\left(\partial_{p+1}\right) .
$$
\end{definition}
The homology groups $H_p(K)$ capture topological features of the shape represented by the simplicial complex $K$. Intuitively its elements represent equivalence classes of $p$ dimensional feature.

For \(p=0\), the boundary map \(\partial_0: C_0 \to 0\) is the zero homomorphism, so \(\ker(\partial_0)=C_0\). The zeroth homology group is therefore
 $$H_0(K) \cong C_0 / \operatorname{im}(\partial_1).$$
This group characterizes the connected components of \(K\), each homology class of $H_0$ corresponds to a distinct connected component of $K$.
Specifically, the rank of \(H_0\) equals the number of connected components.
All vertices within the same connected component are homologous, meaning their differences vanish in the quotient, and thus they represent the same homology class.
Conversely, vertices from different components give rise to distinct classes.
Consequently, if one vertex is chosen from each connected component, the homology class of that vertex serves as a generator for the corresponding component's contribution to \(H_0\).

\begin{definition}[Filtration]
A filtration of a simplicial complex $K$ is a family of subcomplexes $\left\{K_s\right\}_{s\in \mathbb{R}}$ such that
$$
K_t \subseteq K_s \quad \text { for all } t \leq s,
$$
and $K=\bigcup_{s\in \mathbb{R}} K_s .$
In practice, we often use a finite increasing sequence
$${\varnothing} = K_0 \subseteq K_1 \subseteq K_2 \subseteq \cdots \subseteq K_n  = K .
$$
\end{definition}

Given a filtration $\left\{K_s\right\}$, for each dimension $p$ we have a sequence of homology groups with inclusion maps
$$
H_p\left(K_0\right) \rightarrow H_p\left(K_1\right) \rightarrow \cdots \rightarrow H_p\left(K_n\right) .
$$

A $p$-dimensional feature (e.g., a connected component) is born at $s=b$ if it appears in $H_p\left(K_b\right)$ but not in the image from $H_p\left(K_{b-\epsilon}\right)$.
It dies at $s=d$ if it merges with an older feature or becomes a boundary when going from $K_{d-1}$ to $K_d$.


 In this paper, we quantify the peak structures and distribution of 1D graphs using their zeroth homology group \(H_0\).
We introduce a persistent homology framework to analyze the multi-scale evolution of connected components in the graphs of 1D functions.

 For a continuous function \(\gamma(t): [a,b] \to \mathbb{R}\), the 1D functional graph is given by its plot. We compute the persistent homology of \(\gamma\) from its sublevel sets.
Specifically, as the threshold \(s\) increases from \(-\infty\) to \(+\infty\), we track how the connected components of \(\{t: \gamma(t) \le s\}\) appear, disappear, and merge.
Zero-dimensional persistent homology captures the complete lifecycle of these components: their birth time upon first appearance, and their death time upon merging with a component born earlier.
The steps are as follows.

1. Discretization and Simplicial Complex Construction: we discretize the function  $\gamma(t)$ into a simplicial complex $K$.


Let \( \mathbf{T} = \{t_0, t_1, \dots, t_N\} \) be an equidistant partition of the interval \([a, b]\), with \( t_i = a + i \cdot \frac{b-a}{N} \). At each sample point \( t_i \), we associate a Robin coefficient \( \gamma_i = \gamma(t_i) \), which serves as a scalar attribute. Each point–value pair is then represented as a 0-dimensional simplex (vertex) \( v_i \), yielding the vertex set \( V = \{v_0, v_1, \dots, v_N\} \).
To recover a continuous representation of the function over the domain, the discrete vertex data are interpolated using spline functions. This defines smooth functional values along the edges of the resulting graph. The combinatorial structure is completed by connecting each consecutive pair of vertices \( v_i \) and \( v_{i+1} \) to form a 1-dimensional simplex (edge) \( e_{i, i+1} \). The resulting 1-complex \( K = (V, \{e_{i,i+1}\}) \) captures both the geometry of the underlying interval and the variation of the coefficient \( \gamma \) along it.


2.  Filtration Construction and Algebraic Structure:

Based on the simplicial complex $K$, we define a filtration $\{K_s\}_{s \in \mathbb{R}}$ parameterized by the threshold $s$, where $K_s$ consists of all simplices whose filtration values do not exceed $s$:
$$K_s= \{\sigma \in K : \gamma (\sigma) \le s\}.$$
Clearly, for $s \le t$, we have $K_s \subseteq K_t$, giving rise to a sequence of inclusion maps:
$$K_{s_0} \hookrightarrow K_{s_1} \hookrightarrow \cdots \hookrightarrow K_{s_m} = K,$$
where $s_0 < s_1 < \cdots < s_m$ are the distinct filtration values of all simplices.
For each subcomplex $K_s$, we consider its chain complex:
$$\cdots \xrightarrow{\partial_2} C_1(K_s) \xrightarrow{\partial_1} C_0(K_s) \xrightarrow{\partial_0} 0,$$
where $\partial_1$ is the boundary operator mapping each edge to its endpoints: $\partial_1(e_{i,i+1}) = v_{i+1} - v_i$, and $\partial_0$ is the zero homomorphism.
The group $H_0(K_s)$ has $H_0(K_s) \cong C_0(K_s) / \operatorname{im}(\partial_1),$
and its rank equals exactly the number of connected components in the sublevel set $\gamma^{-1}(-\infty, s]$.

The inclusion maps $K_s \hookrightarrow K_t$ induce linear maps between homology groups: $f^{s,t}_0: H_0(K_s) \to H_0(K_t),$
and it forms a persistence module which captures the evolution of connected components across varying filtration parameters :
$$H_0(K_{s_0}) \xrightarrow{f_0} H_0(K_{s_1}) \xrightarrow{f_0} \cdots \xrightarrow{f_0} H_0(K_{s_m}).$$

3. Persistence Pairs

The core utility of persistent homology lies in its ability to track the inception ("birth") and disappearance ("death") of homology classes as the filtration evolves. In the 0-dimensional case (connected components), this process is characterized by three key events:

Birth: A new 0-dimensional homology class $[\alpha] \in H_0(K_s)$ is born at filtration value $s = \gamma_i$ when a vertex $v_i$ first appears in the complex $K_s$. This signifies the emergence of a new disjoint connected component.

Merge (The Elder Rule): When an edge $e_{i,i+1}$ is introduced at the threshold $s = \max\{\gamma_i, \gamma_{i+1}\}$, it may bridge two previously disjoint components. If these components merge, their relative significance is determined by the Elder Rule:

a.	The homology class with the more recent birth time (the "younger" class) is considered to die, being subsumed into the class with the earlier birth time (the "elder" class).

b.	This event defines a persistence pair $(s_{birth}, s_{death})$, where $s_{death}$ corresponds to the filtration value of the merging edge.

Persistence: The lifespan of a homology class is quantified as its persistence, defined by the interval length $s_{death} - s_{birth}$. This metric serves as a proxy for the topological significance of a feature across scales; high persistence typically indicates robust structural features, while low persistence often corresponds to noise.

The persistent distance is defined as:
\begin{equation}
\lVert \gamma \rVert_{\text{per}} = \lVert \gamma|_{\mathbf{T}} \rVert:= \sum_{(t_k, t_l) \in P_1} \vert \gamma(t_l) - \gamma(t_k) \vert + \sum_{(t_k, t_l) \in P_2} \vert \gamma(t_l) - \gamma(t_k) \vert,
\end{equation}
where \( P_1 \) denotes the set of persistence pairs of \( \gamma \) (i.e., pairs of filtration parameters corresponding to the birth and death of connected components in the sublevel set filtration of \( \gamma \)), and \( P_2 \) denotes the set of persistence pairs of \( -\gamma \) (i.e., the superlevel set filtration of \( \gamma \)). 


Following the work of Zheng\cite{Zheng2015}, we distinguish between two types of persistence pairs. 
A single persistence pair appears only in either the sublevel set filtration ($P_1$) or the superlevel set filtration ($P_2$), but not in both. 
A double persistence pair belongs to $P_1 \cap P_2$, i.e., it is present in both filtrations. 
The set of all persistence pairs is $P(u) = P_1(u) \cup P_2(u)$. 
We further define 
\[
S(u) := P(u) \setminus \bigl(P_1(u) \cap P_2(u)\bigr),
\]
which consists of all single persistence pairs together with the global minimum–maximum pair (treated as a single pair).
For each double persistence pair $(t_j, \tilde{t}_j)$, its order $k = k(t_j, \tilde{t}_j)$ is defined as the number of larger persistence intervals that strictly contain $[t_j, \tilde{t}_j]$ in the unique chain of interlacing intervals  in \cite{Zheng2015}). 
In particular, all pairs in $S(u)$ are assigned order $0$. 
This notion of order captures the nesting depth of topological features, which is distinct from the notion of persistence measured by $|\tilde{t}_j - t_j|$.


The persistent distance thus sums the functional value distances of all persistent pairs in \( P_1 \) and \( P_2 \). Smaller distances \( \vert \gamma(t) - \gamma(\tilde{t}) \vert \) corresponding to short persistent pairs \( (t, \tilde{t}) \) reflect noise-like oscillatory behavior, while larger distances characterize significant features of \( \gamma \).

\subsection{Bayesian inference fundamentals }

We consider the forward model
\[
Y^{\dagger} = F(\gamma),
\]
where \( F(\gamma)(t) := u(0, t) \) is the solution to $\eqref{eqn1}$ at the boundary $x=0$. 
Observations are collected at times \( t_1, \dots, t_M \) and are assumed to satisfy
\begin{equation}\label{eqndata}
Y_n = F(\gamma)(t_n) + \eta_n,\quad n=1,\dots,M,
\end{equation}
where \( \eta = (\eta_1, \dots, \eta_M)^\top \) is a zero-mean Gaussian random vector with symmetric positive definite covariance matrix \( \Sigma \).
Denote $Y=[Y_1, \cdots, Y_M]^\top $.
We aim to estimate the posterior distribution  of $\gamma$, denoted by $\mu_{\text{post}}$,  given the observed data ${Y}$. This posterior provides a complete solution to the inverse problem, delivering both a parameter estimate and principled uncertainty quantification.

For inverse problems, prior information is essential as it provides the necessary constraints or regularization to render the ill-posed problem tractable. This allows the incorporation of known physical properties, bounds, or structural assumptions regarding the unknown parameter into the inference process. Here, the prior knowledge about $\gamma$ is formally encoded through a prior distribution $\mu_{\text{pr}}$, which is defined over an appropriate Hilbert space.

Bayes' formula in this infinite-dimensional setting connects the prior and posterior distributions via the Radon-Nikodym derivative, provided that the posterior measure \(\mu_{\text{post}}\) is absolutely continuous with respect to the prior measure \(\mu_{\text{pr}}\):
\[
\frac{d\mu_{\text{post}}}{d\mu_{\text{pr}}}(\gamma) \propto \exp\!\bigl( -\Phi(\gamma; Y) \bigr),
\]
where \(\Phi(\gamma; Y)\) denotes the negative log-likelihood (or data misfit) functional

\[
\Phi(\gamma; Y) := \frac{1}{2} \| F(\gamma) - Y \|_\Sigma^2 = \frac{1}{2} \| \Sigma^{-1/2} (F(\gamma) - Y) \|_2^2.
\]
The most widely used prior in Bayesian inverse problems is the Gaussian prior, i.e., $\mu_{\text{pr}} = \mu_0$, where $\mu_0 = N(0, C_0)$. In practical terms, the Gaussian prior reflects our belief that the unknown parameter $\gamma$ has a mean of 0 before any data is observed. The covariance $C_0$ represents our assumption about the variability or uncertainty in $\gamma$.

In this work, we adopt a Gaussian process prior $\gamma \sim \mathcal{GP}(0, k)$ with zero mean and the squared exponential covariance function
\[
k(x,y) = \exp\left(-\frac{|x-y|^2}{2l^2}\right),
\]
where $l>0$ is a length-scale parameter. The corresponding covariance operator $C_0$ is defined by
\[
(C_0 u)(x) = \int_{\Omega} k(x,y) \, u(y) \, dy, \qquad u \in L^2(\Omega).
\]
This operator is self-adjoint, positive definite, and trace-class on $L^2(\Omega)$ under suitable conditions.

%
%
%


In order to capture non-smooth features, we add extra constraints to the Gaussian prior and form a hybrid prior. The idea follows Yao et al. \cite{Yao2016}: integrate modeling and constraints of different types within one framework to increase prior flexibility. This hybrid strategy improves the representation of complex systems, especially when both uncertainty and structural information are present.

In \cite{Yao2016}, the authors adopt hybrid priors and suppose
\[
\frac{d\mu_{\text{pr}}}{d\mu_0}(\gamma) \propto \exp\left(-R(\gamma)\right),
\]
instead of simply letting \(\mu_{\text{pr}} = \mu_0\), where \(R(\gamma)\) represents additional prior information on \(\gamma\). Thus, the Radon-Nikodym derivative of the posterior measure \(\mu_{\text{post}}\) with respect to \(\mu_0\) can be expressed as:
\[
\frac{d\mu_{\text{post}}}{d\mu_0}(\gamma) \propto \exp\left(-\Phi(\gamma; Y) - R(\gamma)\right).
\]
When \(R(\gamma) = 0\), this reduces to the form under the standard Gaussian prior.

Following \cite{Yao2016}, a hybrid TV-Gaussian prior can be constructed by choosing $R(\gamma)$ to be the total variation functional $R_{\text{TV}}$, yielding:
\begin{equation}\label{eqntv}
\frac{d\mu_{\text{pr}}}{d\mu_0}(\gamma) \propto \exp\left(-R_{\text{TV}}(\gamma)\right).
\end{equation}
In the numerical implementation, the domain is discretized into \(N_t+1\) grid points, and the TV regularization term can be approximated as:
\begin{equation}\label{termtv}
R_{\text{TV}} = \lambda\sum_{i=1}^{N_t} |\gamma_{i+1} - \gamma_i|.
\end{equation}

For the one-dimensional case, the PH-Gaussian prior distribution is defined as \cite{Deng2025}:
\begin{equation}\label{eqnph}
\frac{d\mu_{\text{pr}}}{d\mu_0}(\gamma) \propto \exp\left( -R_{\text{PH}}(\gamma)\right),
\end{equation}
The regularization term \( R_{\text{PH}}(\gamma)\) is expressed as:
\begin{equation}\label{termph}
R_{\text{PH}}(\gamma) = \lambda\sum_{(t_j, \tilde{t}_j) \in P(\gamma)} \alpha_j(\gamma) \vert \gamma(t_j) - \gamma(\tilde{t}_j) \vert,
\end{equation}
where the weights \(\alpha_j = \alpha_j(\gamma) = \alpha(\gamma, t_j, \tilde{t}_j)\) depend on the persistence \(\vert \gamma(t_j) - \gamma(\tilde{t}_j) \vert\): for smaller distances \(\vert \gamma(t_j) - \gamma(\tilde{t}_j) \vert\), \(\alpha_j\) should take larger values; for larger distances \(\vert \gamma(t_j) - \gamma(\tilde{t}_j) \vert\), \(\alpha_j\) should take smaller values. 

By appropriately selecting \(\alpha_j\), the reconstructed function \(\gamma\) can effectively retain the key features of the original \(\mathbf{y} = (\gamma(t_j))_{j=0}^N\). Therefore, we need to consider not only the local behavior of the function on the interval \([t_l, t_{l+1}]\), but also incorporate the structural information of the corresponding persistence pairs. If prior knowledge about the hierarchical structure of the original signal (such as the hierarchy level of the chains) is available, it becomes possible to more accurately determine which persistence pairs represent significant features. Noise typically corresponds to high-order persistence pairs, while low-order pairs reflect important structures. Based on the hierarchical information of the chains, for any persistence pair \((t_j, \tilde{t}_j) \in P(\gamma)\), the weight can be set as:
\begin{align}
\alpha_j(\gamma) = {(k_j + 1)} \cdot \frac{1}{1 + \eta \vert \gamma(\tilde{t}_j) - \gamma(t_j) \vert},
\end{align}
where  \(k_j = k(t_j, \tilde{t}_j)\) denotes the hierarchy level of the persistence pair \((t_j, \tilde{t}_j)\) within the chain, and \(\eta > 0\) is a parameter that adjusts the sensitivity to functional value differences.

%
%

\subsection{Hierarchical Bayesian models }

In the Bayesian inversion of the Robin coefficient $\gamma$, the regularization parameter $\lambda$ is pivotal in governing the smoothness of the reconstructed profile. The prior distribution, which penalizes high-frequency oscillations, is scaled by $\lambda$. While a larger $\lambda$ enforces structural regularity and suppresses noise, a smaller $\lambda$ prioritizes data fidelity at the risk of overfitting.

In the absence of a universal criterion or closed-form solution for an optimal $\lambda$, its selection introduces significant epistemic uncertainty. To mitigate this, we adopt a hierarchical Bayesian approach, treating $\lambda$ as an unknown hyperparameter. By assigning $\lambda$ a Gamma prior distribution, we enable its joint inference with $\gamma$ from the observations $Y$. This allows for an autonomous, data-driven balance between observational accuracy and prior smoothness constraints. The resulting joint posterior density is given by:
$$ p(\gamma, \lambda | Y) \propto p(Y | \gamma) p(\gamma | \lambda) p(\lambda). $$

Regarding the hyperparameter $\lambda$, a conjugate prior is a conventional choice to ensure computational efficiency. In equations \eqref{eqntv} and \eqref{eqnph}, we assign a Gamma prior to $\lambda$, defined as:
\[p(\lambda) \propto \lambda^{m_1 - 1} \mathrm{e}^{-m_2 \lambda}, \] where $m_1$ and $m_2$ are the shape and rate parameters, respectively. This conjugate formulation is particularly advantageous as it allows the conditional posterior of $\lambda$ to be updated in a closed-form, facilitating the use of efficient sampling algorithms such as the Gibbs sampler.

%

%

%

Combining the above prior structure with the prior for $\lambda$, the hierarchical Bayesian posterior distribution can be computed as:
\[
p(\gamma, \lambda \mid Y) \propto \exp\left(-\frac{1}{2\sigma^2}\|\mathbf{F}(\gamma) - Y\|_2^2\right) \exp\left(-\frac{1}{2}\|\gamma\|_E^2\right)\exp\left(- R(\gamma)\right) \lambda^{m_1 - 1} \mathrm{e}^{-m_2 \lambda}
\]
and the full conditional distribution$p(\lambda \mid \gamma,Y)$ can be given as
\[
p(\lambda \mid \gamma,Y) \propto \lambda^{m_1 - 1}   \mathrm{exp}(-(m_2 \lambda+R(\gamma)))
\]

\subsection{Numerical exploration of the posterior state space }

This paper employs the preconditioned Crank–Nicolson (pCN) and Gibbs algorithm to sample from the posterior distribution $p(\gamma,\lambda|Y)$, obtaining representative samples of the posterior parameters for further statistical analysis of the state space.

In the numerical experiments, the synthetic observational data are generated according to \eqref{eqndata} by adding relative Gaussian noise to the numerical solution of the forward map $F(\gamma)$ at the discrete grid points, with noise covariance $\Sigma = \epsilon \, \|F(\gamma)\| \, I$, where $\epsilon$ denotes the relative noise level.

\begin{algorithm}
\caption{}
\label{alg:myalg}
\begin{algorithmic}[1]
\State \textbf{Initialization:} Set sample size \( N_s \). Take \( \gamma^{(0)}  \), \( \lambda^{(0)} \). Denote
$\varphi^{(0)} := \Phi(\gamma^{(0)}; Y) + \lambda R(\gamma^{(0)})+ \frac{1}{2}\|\gamma^{(0)}\|_E^2- \log p(\lambda^{(0)})$

\State Propose a candidate sample $\hat{\gamma}^{(n)} = \sqrt{1 - \beta^2} \gamma^{(n-1)} + \beta \xi$, where $\xi \sim \mathcal{N}(0, C_0);$

\State Denote
$
\hat{\varphi}^{(n)} := \Phi(\hat{\gamma}^{(n)}; Y) +\lambda R(\hat{\gamma}^{(n)})+ \frac{1}{2}\|\hat\gamma^{(n)}\|_E^2- \log p({\lambda}^{(n-1)})$.

\State Compute the acceptance probability \( \alpha = \min\left\{1, \exp\left[-\hat{\varphi}^{(n)} + \varphi^{(n-1)}\right]\right\} \).

\State Update \( \gamma^{(n)}\): If \( \alpha > \theta \), set \( \gamma^{(n)} = \hat{\gamma}^{(n)} \); else set \( \gamma^{(n)} = \gamma^{(n-1)} \), where \( \theta\sim U(0, 1) \).

\State Update $\lambda^{(n)}$ via Gibbs sampling: \ $\lambda^{(n)} \sim \Gamma\left( m_1,\ R(\gamma) + m_2 \right)$

\State While \( n < N_s \), repeat Steps 2–6.
\end{algorithmic}
\end{algorithm}

In the aforementioned algorithm, the regularization term ${R(\gamma)}$ is given by \eqref{termtv} for the TV–Gaussian prior and and \eqref{termph} for the PH–Gaussian prior. A burn-in phase is incorporated in the sampling procedure. We discard the first 30$\%$ of the initially generated samples are discarded as the burn-in period to minimize the influence of the initial distribution on stationary sampling. Finally, the mean of all post-burn-in samples is adopted as the posterior estimate of the parameter.

\section{Numerical experiments}\label{stest}

In this section, we consider three one-dimensional examples to assess the effectiveness of the PH-Gaussian prior for identifying the Robin coefficient. We conduct the inversion using three different priors: a Gaussian prior, a TV–Gaussian prior and a PH–Gaussian prior and compare the numerical results obtained under each prior. The spatial domain  $\Omega$ for the numerical examples is set to $(0,1)$ and the final time is fixed at $T = 1$. Temperature is measured on $\Gamma_c = \{x = 0\}$, while the time-dependent Robin coefficient $\gamma(t)$ on $\Gamma_i = \{x = 1\}$ is to be estimated. The spatial domain $\Omega$ is discretized into 100 elements and the number of time steps is 200. For all experiments, the number of samples is set to $N_s = 10^5$. The chosen hyperparameters are \( m_1 = 50 \) and \( m_2 = 0.1 \) for  Example~\ref{example1} and Example~\ref{example2}, while \( m_1 = 200 \) and \( m_2 = 0.1 \) for Example~\ref{example3}. In all examples, the observations $Y$ are generated as \eqref{eqndata}.

To evaluate the accuracy of the numerical inversion results $\gamma(t)$, this paper adopts the following relative error metric:
\[
e_r:=\operatorname{error}(\gamma) = \frac{\sqrt{\sum_{i=1}^{m} (\gamma_i - \hat{\gamma}_i)^2}}{\sqrt{\sum_{i=1}^{m} (\gamma_i)^2}}
\]
where $\gamma_i$ and $\hat{\gamma}_i$ represent the exact value and the inverted value at the discrete node $t_i$, respectively.

\begin{example}
\label{example1}
The conductivity $\alpha(x)$ is set as $\alpha = 1$, and the exact Robin coefficient $\gamma^{\dagger}(t)$ is given by $\gamma^{\dagger}(t) = 2 + \sin(2\pi t)$. The boundary conditions, source term, and initial condition are specified such that the analytical solution to the direct
problem \eqref{eqn1} is given by $u(x,t) = e^{2x} \sin(t)$.
\end{example}

\begin{figure}[htbp]
    \centering  

    \begin{subfigure}[b]{0.48\textwidth}  
        \centering
        \includegraphics[width=\textwidth]{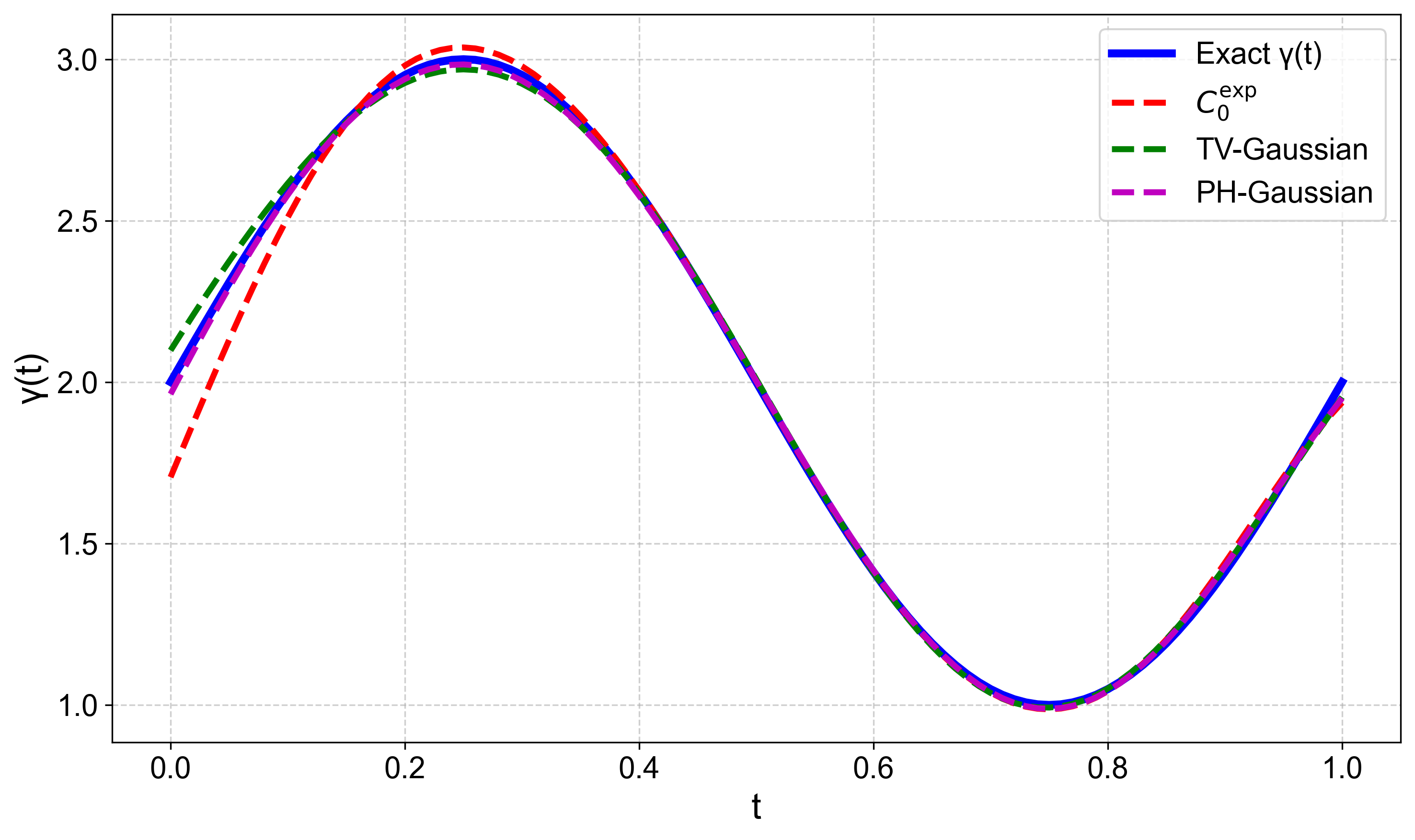}  
        \caption{$\epsilon=1\%$}  
        \label{eg1fig:sub1}  
    \end{subfigure}
    \hspace{0.01\textwidth}  
    \begin{subfigure}[b]{0.48\textwidth}
        \centering
        \includegraphics[width=\textwidth]{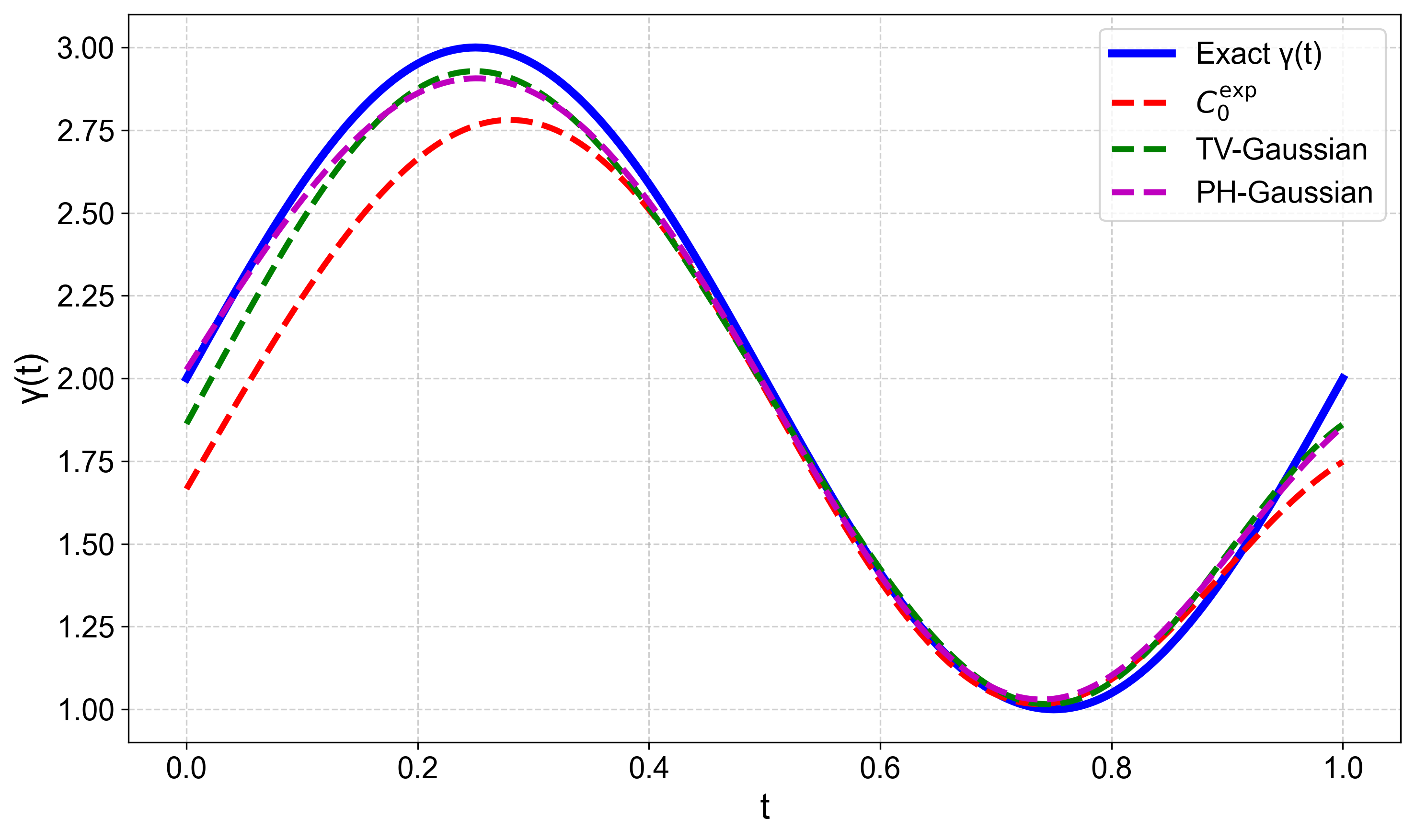}
        \caption{$\epsilon=5\%$}
        \label{eg1fig:sub2}
    \end{subfigure}

    \caption{The numerical results for Example~\ref{example1} with (a) 1$\%$ and (b) 5$\%$ noise added into the data.}
    \label{fig:main1}  
\end{figure}

In this example, we investigate the reconstruction of a smooth temporal Robin coefficient $\gamma(t)$ from noisy measurements. 
The performance of the proposed PH-Gaussian prior is evaluated against the standard Gaussian prior and the TV-Gaussian prior under three distinct noise levels ($\epsilon = 0.01, 0.03, 0.05$). 
The relative errors ($e_r$) summarized in Table \ref{table1} demonstrate that the PH-Gaussian prior consistently outperforms the other two methods across all noise levels. 
Specifically, at the $\epsilon = 0.01$ level, the PH-Gaussian prior yields an error of $0.0064$, representing a significant reduction compared to the Gaussian ($0.0315$) and TV-Gaussian ($0.0127$) priors. 
This indicates that by integrating persistent homology, the prior more effectively captures the global topological features and inherent smoothness of the target coefficient. 
As the noise level increases to $5\%$, the error for the PH-Gaussian prior grows more gracefully (from $0.0064$ to $0.0261$) than the sharper increase observed with the Gaussian prior. 
Visual inspection of Fig. \ref{fig:main1} confirms that the PH-Gaussian estimate tracks the exact solution with high fidelity, particularly at the peaks where other methods tend to undershoot.

\begin{figure}[htbp]
    \centering  

    \begin{subfigure}[b]{0.45\textwidth}  
        \centering
        \includegraphics[width=\linewidth]{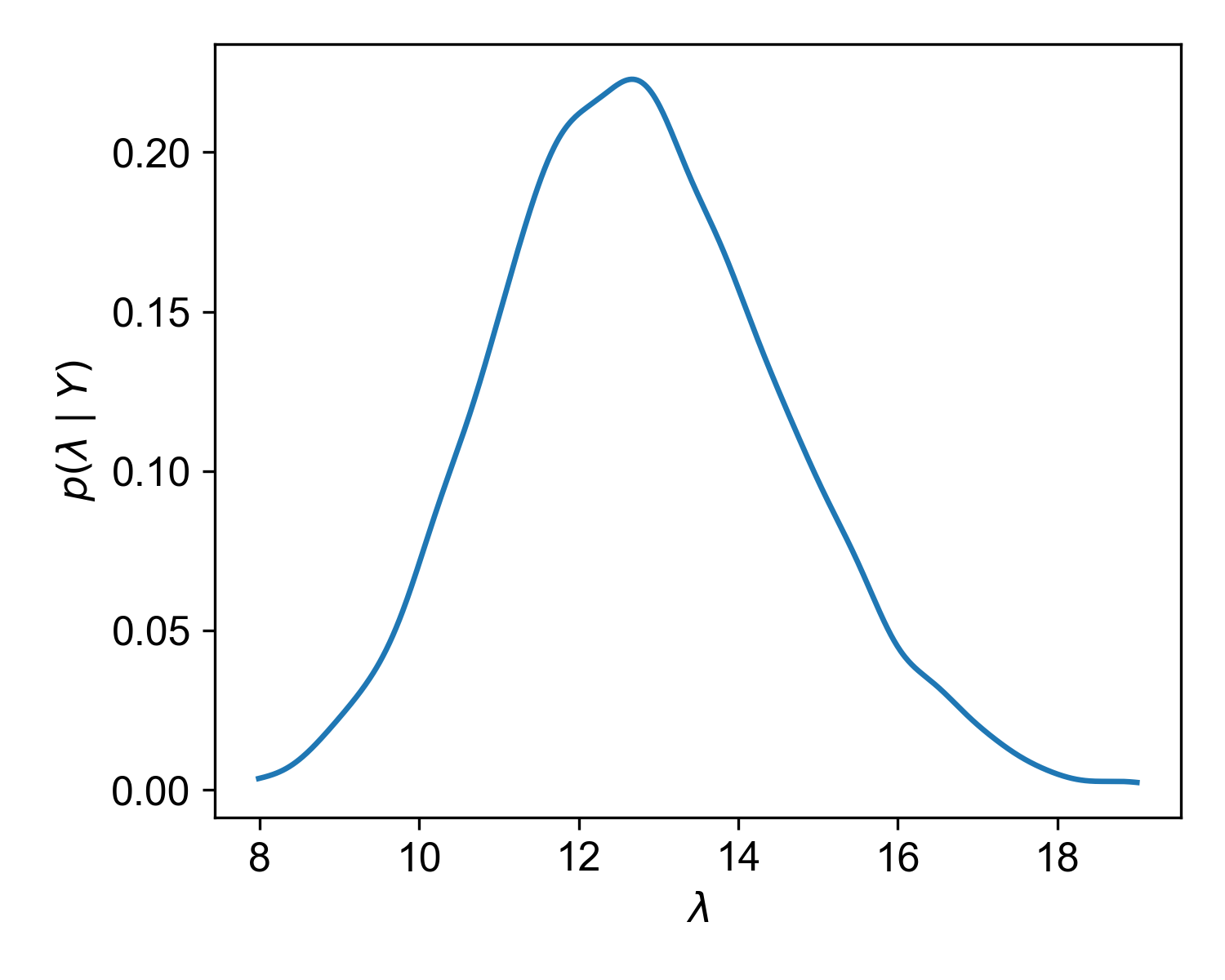}  
        \caption{The posterior density of TV–Gaussian}  
        \label{eg1fig:sub1hec}  
    \end{subfigure}
    \hfill
    \begin{subfigure}[b]{0.45\textwidth}
        \centering
        \includegraphics[width=\linewidth]{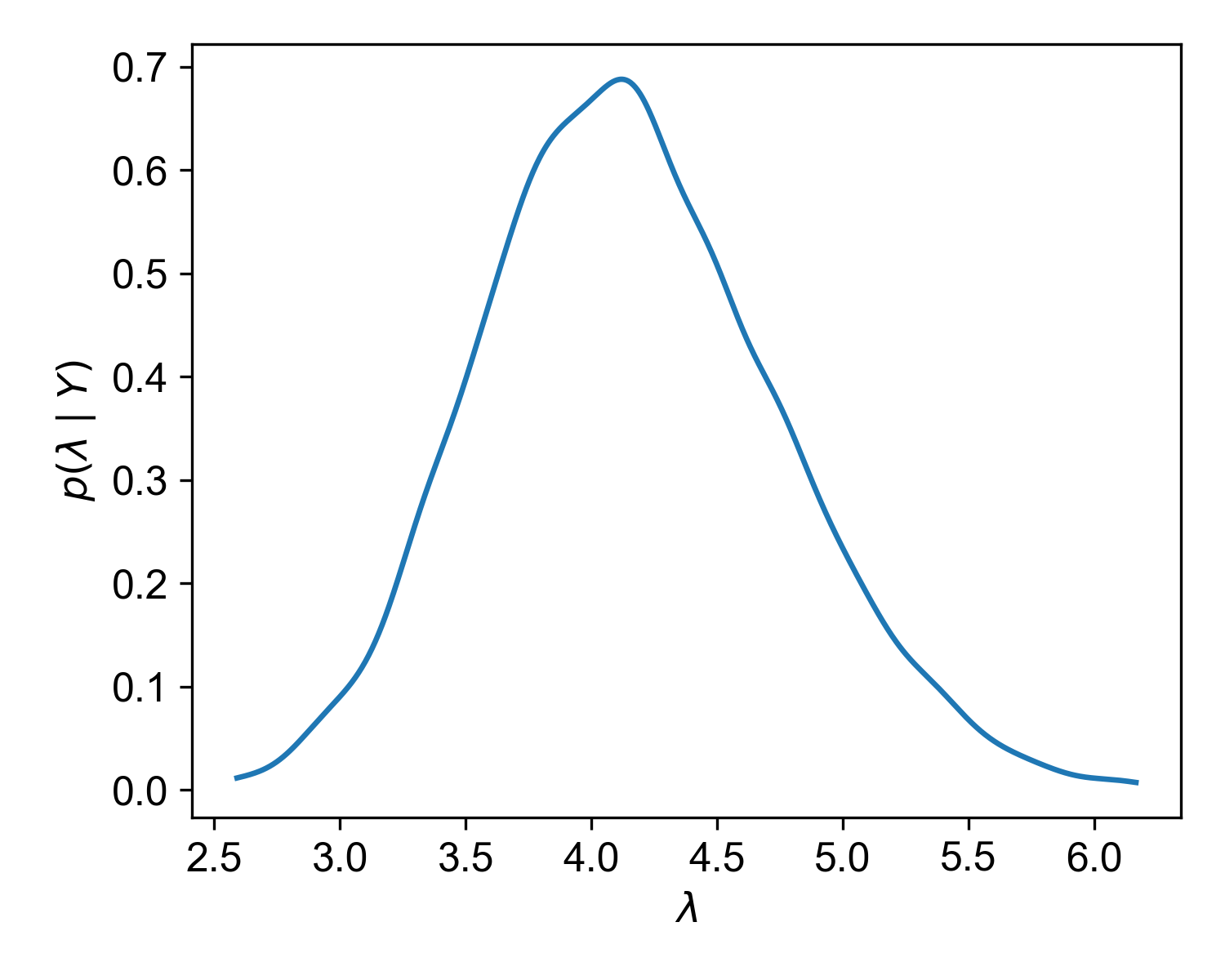}
        \caption{The posterior density of PH–Gaussian}
        \label{eg1fig:sub2hec}
    \end{subfigure}
\hfill
    \caption{The posterior density of the scaling parameter \(\lambda\) for Example~\ref{example1} with 1$\%$ noise added into the data.}
    \label{fig:main2}  
\end{figure}

The posterior density plots for $\lambda$ (Fig. \ref{fig:main2}) demonstrate the efficacy of the hierarchical framework. The unimodal distribution of $p(\lambda | Y)$ indicates that the model successfully identifies the optimal regularization strength without manual tuning. As noise rises, the estimate $\hat{\lambda}$ for the PH-Gaussian prior increases from $4.16$ to $4.61$ (Table \ref{table1}), reflecting an adaptive mechanism that strengthens the prior constraint to counter data uncertainty. Additionally, the scale difference in $\hat{\lambda}$ between TV-Gaussian ($\approx 12$) and PH-Gaussian ($\approx 4$) priors highlights the distinct regularization mechanisms of topological versus gradient-based constraints.

\begin{figure}[htbp]
    \centering  

    \begin{subfigure}[b]{0.45\textwidth}  
        \centering
        \includegraphics[width=\linewidth]{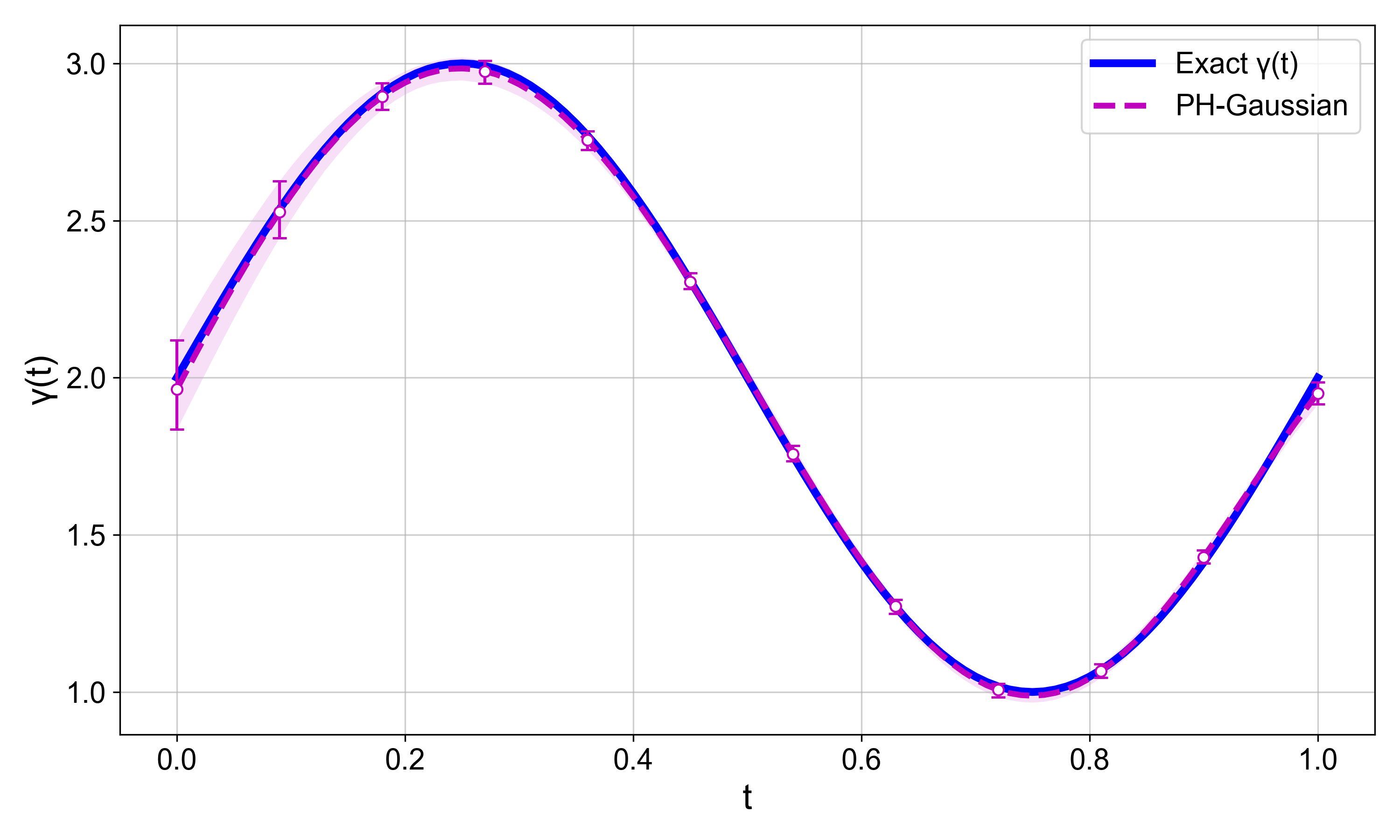}  
        \caption{ $\epsilon=1\%$ }  
        \label{eg1fig:sub1reb}  
    \end{subfigure}
\hfill
    \begin{subfigure}[b]{0.45\textwidth}
        \centering
        \includegraphics[width=\linewidth]{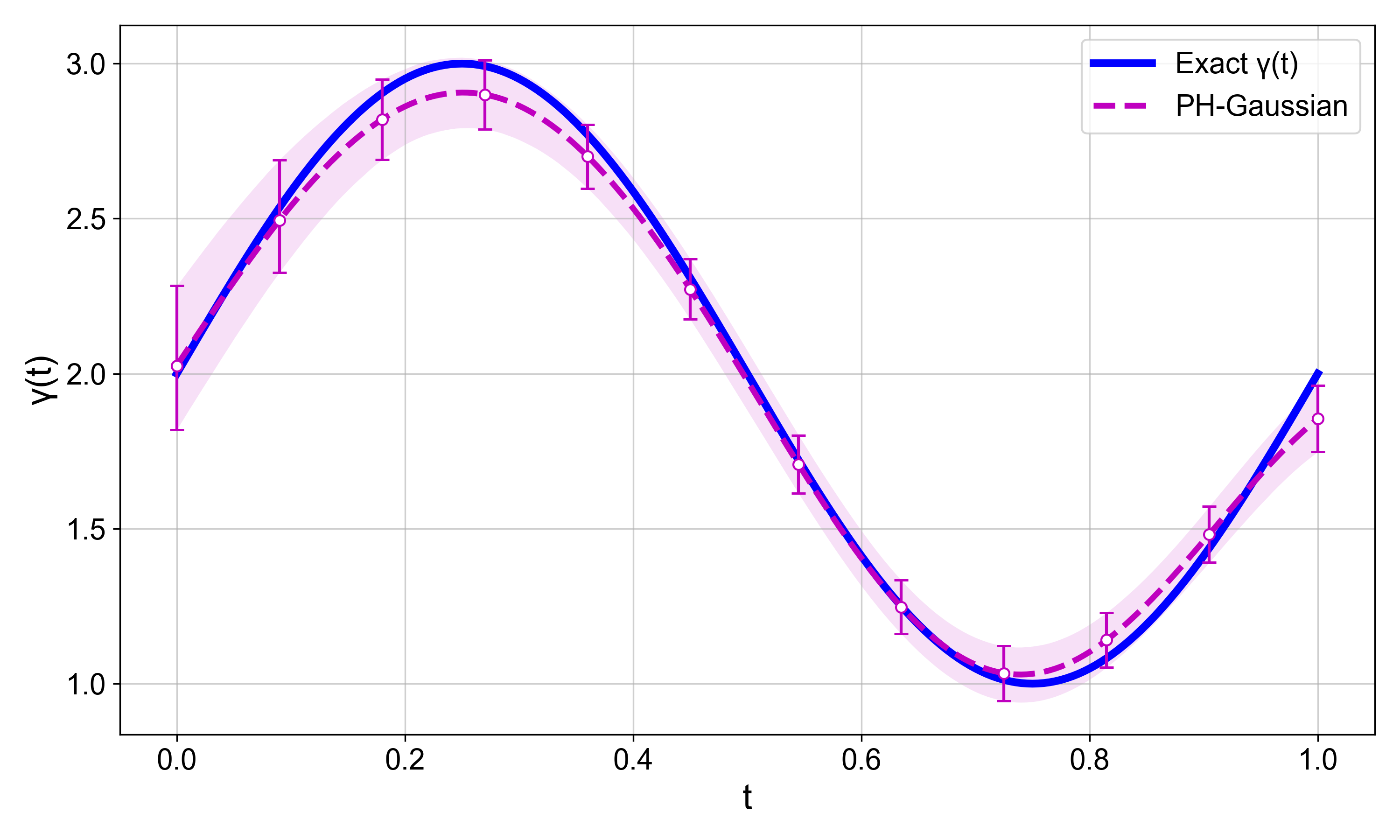}
        \caption{$\epsilon=5\%$  }
        \label{eg1fig:sub2reb}
    \end{subfigure}

\caption{Posterior mean estimates and uncertainty quantification for Example~\ref{example1} at noise levels $\epsilon = 0.01$ and $\epsilon = 0.05$}
    \label{fig:main3}  
\end{figure}

The posterior credible intervals in Fig. \ref{fig:main3} provide an uncertainty quantification analysis of the reconstruction. As the noise level increases from $1\%$ to $5\%$, a clear expansion of the uncertainty bounds is observed, particularly near the boundaries ($t=0, t=1$) and regions of rapid variation. This widening reflects the diminished information content in the noisy data and the resulting increase in localized uncertainty. By integrating the PH-Gaussian prior within a hierarchical framework, the method yields a statistical measure of the estimation risks, allowing for a more nuanced interpretation of the inverted Robin coefficients under different noise conditions.

\begin{table}[htbp]
    \caption{Numerical results of Example~\ref{example1} with different noise levels}
    \centering
    \begin{tabular}{c|cc|cc|cc}
        \toprule
        $\epsilon$ & \multicolumn{2}{c}{Gaussian Prior} & \multicolumn{2}{c}{TV-Gaussian Prior} & \multicolumn{2}{c}{PH-Gaussian Prior} \\
        \cmidrule(lr){2-3} \cmidrule(lr){4-5} \cmidrule(lr){6-7}
                   & $\hat{\lambda}$ & $e_r$ & $\hat{\lambda}$ & $e_r$ & $\hat{\lambda}$ & $e_r$ \\
        \midrule
        0.01 & / & 0.0315 & 12.61 & 0.0127 & 4.16 & 0.0064 \\
        0.03 & / & 0.0507 & 12.19 & 0.0190 & 4.28 & 0.0124 \\
        0.05 & / & 0.0840 & 12.72 & 0.0322 & 4.61 & 0.0261 \\
        \bottomrule
    \end{tabular}
    \label{table1}
\end{table}

\FloatBarrier

\begin{example}
\label{example2}
The conductivity $\alpha(x)$ is set as $\alpha(x) = 1$, and the exact Robin coefficient 
$\gamma^{\dagger}(t)$ is given by 
\[
\gamma(t) =
\begin{cases}
2 t, & 0 \leq t \leq \frac{1}{2}, \\
2 (1 - t), & \frac{1}{2} < t \leq 1.
\end{cases}.
\]
 The boundary conditions, source term, and initial condition are specified such that the analytical solution to the direct
problem \eqref{eqn1} is given by $u(x,t) = (x+1)(t+1)$.
\end{example}

In Example \ref{example2}, we consider a peak-shaped Robin coefficient $\gamma(t)$ with a non-differentiable tip at $t=0.5$. This case evaluates the PH-Gaussian prior's ability to recover sharp geometric features under noise.

As shown in Table \ref{table2}, the PH-Gaussian prior consistently achieves the lowest relative errors across all noise levels. At $\epsilon = 0.01$, its error ($0.0248$) is significantly lower than that of the Gaussian ($0.0857$) and TV-Gaussian ($0.0345$) priors. Notably, the PH-Gaussian approach avoids the staircase typical of TV regularization and the over-smoothing at the tip seen in standard Gaussian methods. Even at $5\%$ noise, it remains robust with an error of only $0.0673$.

\begin{figure}[htbp]
    \centering  

    \begin{subfigure}[b]{0.45\textwidth}  
        \centering
        \includegraphics[width=\linewidth]{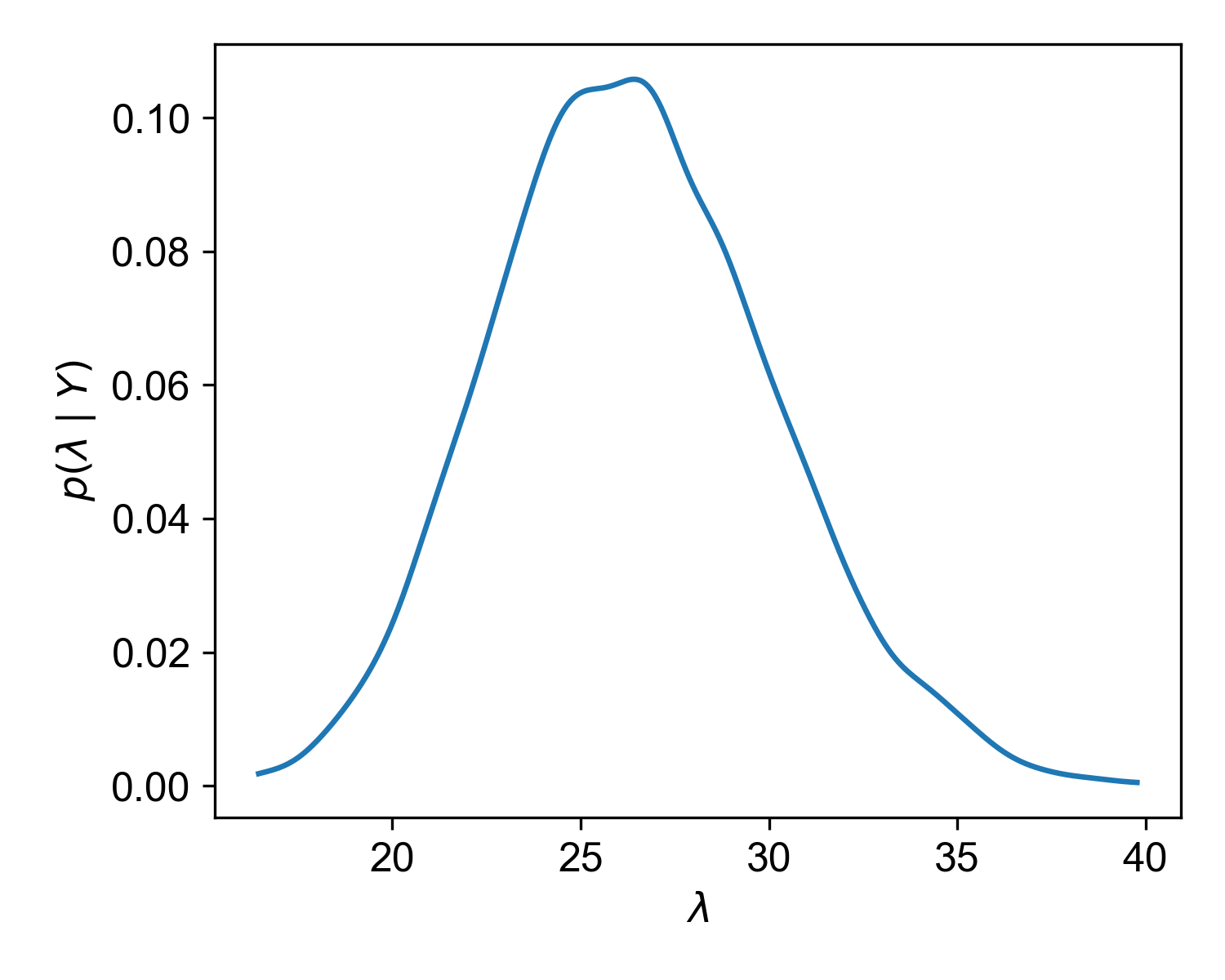}  
        \caption{The posterior density of TV–Gaussian}  
        \label{fig:sub1}  
    \end{subfigure}
    \hfill  
    \begin{subfigure}[b]{0.45\textwidth}
        \centering
        \includegraphics[width=\linewidth]{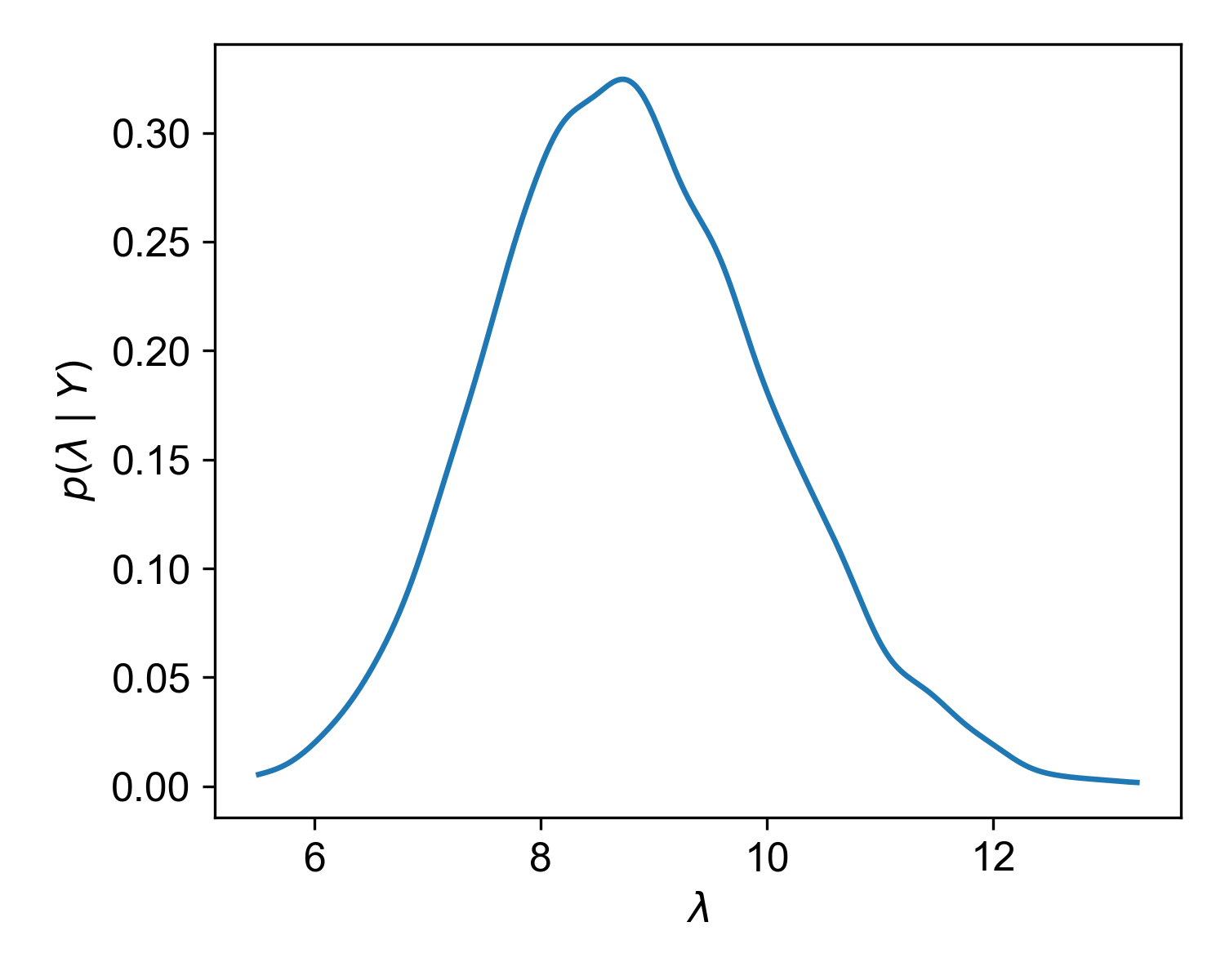}
        \caption{The posterior density of PH–Gaussian}
        \label{fig:sub2}
    \end{subfigure}
    \caption{The posterior density of the scaling parameter \(\lambda\) for Example~\ref{example2} with 1$\%$ noise added into the data.}
    \label{fig:main5}  
\end{figure}

The unimodal posterior distributions of $\lambda$ in Fig. \ref{fig:main5} confirm the efficiency of the hierarchical framework. The adaptive increase of $\hat{\lambda}$ from $8.82$ to $10.29$ as noise rises demonstrates the model's ability to automatically strengthen geometric constraints to counter data uncertainty.

\begin{figure}[htbp]
    \centering  
    \begin{subfigure}[b]{0.45\textwidth}  
        \centering
        \includegraphics[width=\textwidth]{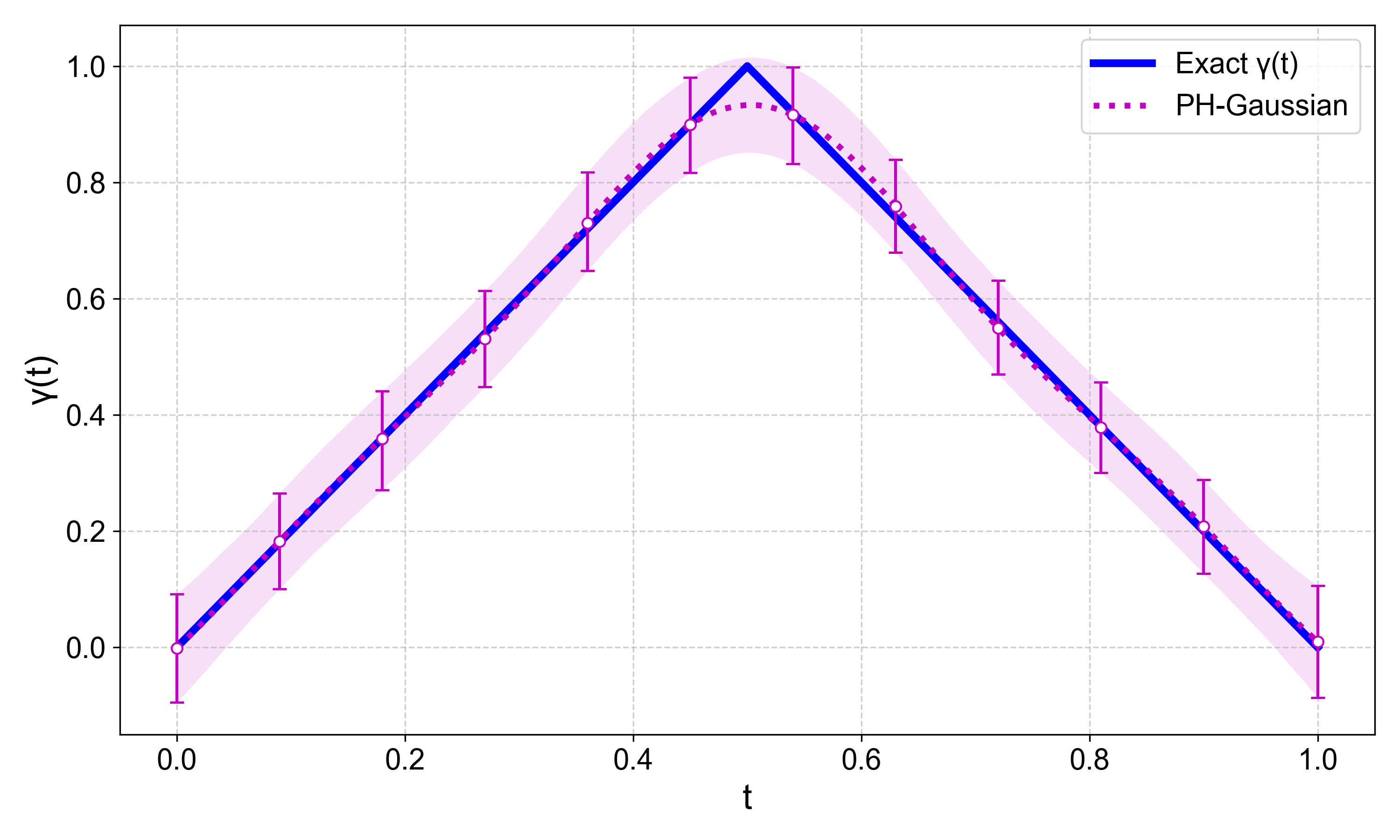} 
        \caption{$\epsilon=1\%$}  
        \label{fig:sub1}  
    \end{subfigure}
    \hfill
    \begin{subfigure}[b]{0.45\textwidth}
        \centering
        \includegraphics[width=\textwidth]{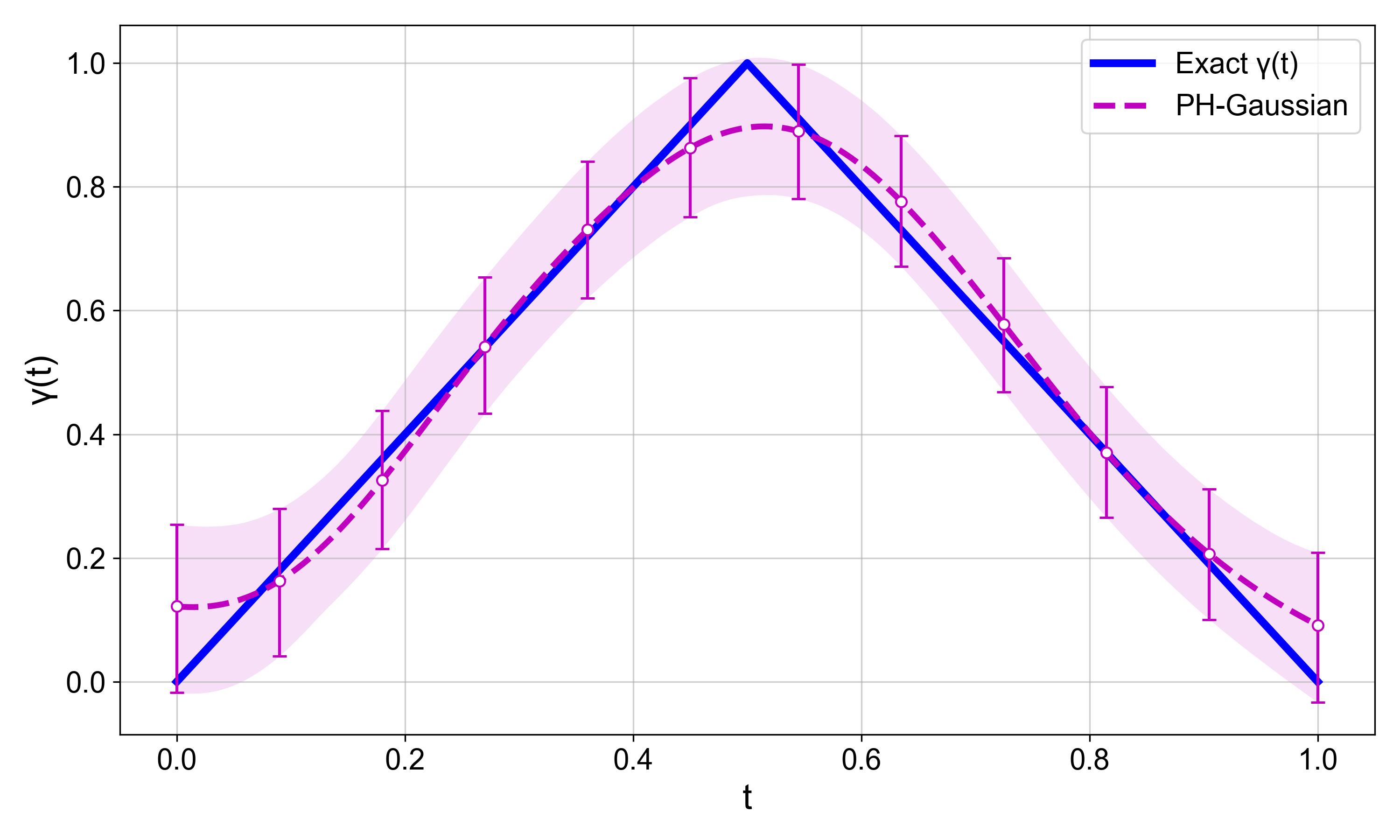} 
        \caption{$\epsilon=5\%$}  
        \label{fig:sub2}
    \end{subfigure}

    \caption{The numerical results for Example~\ref{example2} with (a) 1$\%$ and (b) 5$\%$ noise added into the data.}
    \label{fig:main6}  
\end{figure}

 Fig.~\ref{fig:main6}  shows that the credible intervals expand with increasing noise, particularly around the tip at $t=0.5$. This widening reflects the heightened localized uncertainty in capturing singular features. 

\begin{figure}[htbp]
    \centering  

    \begin{subfigure}[b]{0.48\textwidth}  
        \centering
        \includegraphics[width=\textwidth]{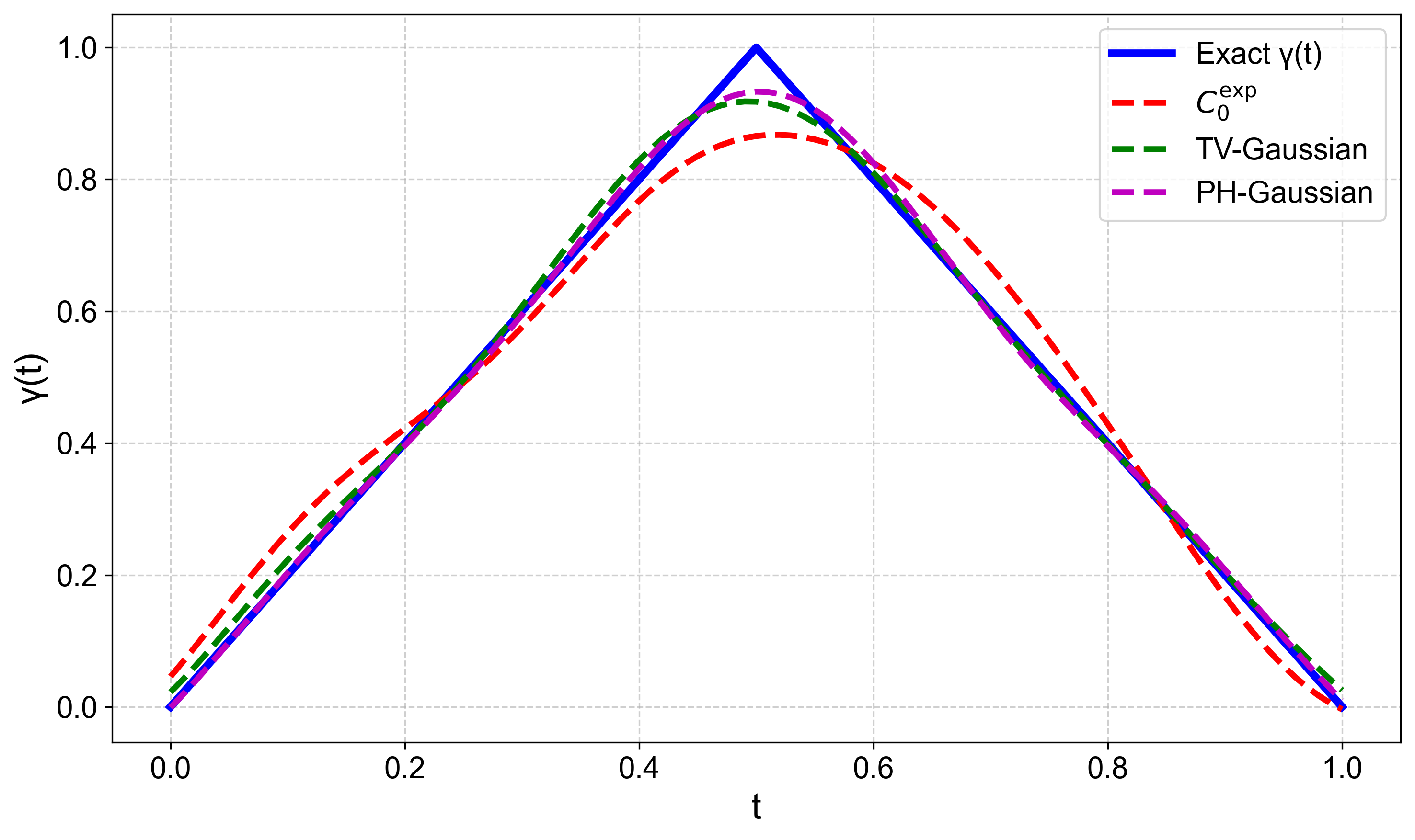}  
        \caption{$\epsilon=1\%$}  
        \label{fig:sub1}  
    \end{subfigure}
    \hspace{0.01\textwidth}  
    \begin{subfigure}[b]{0.48\textwidth}
        \centering
        \includegraphics[width=\textwidth]{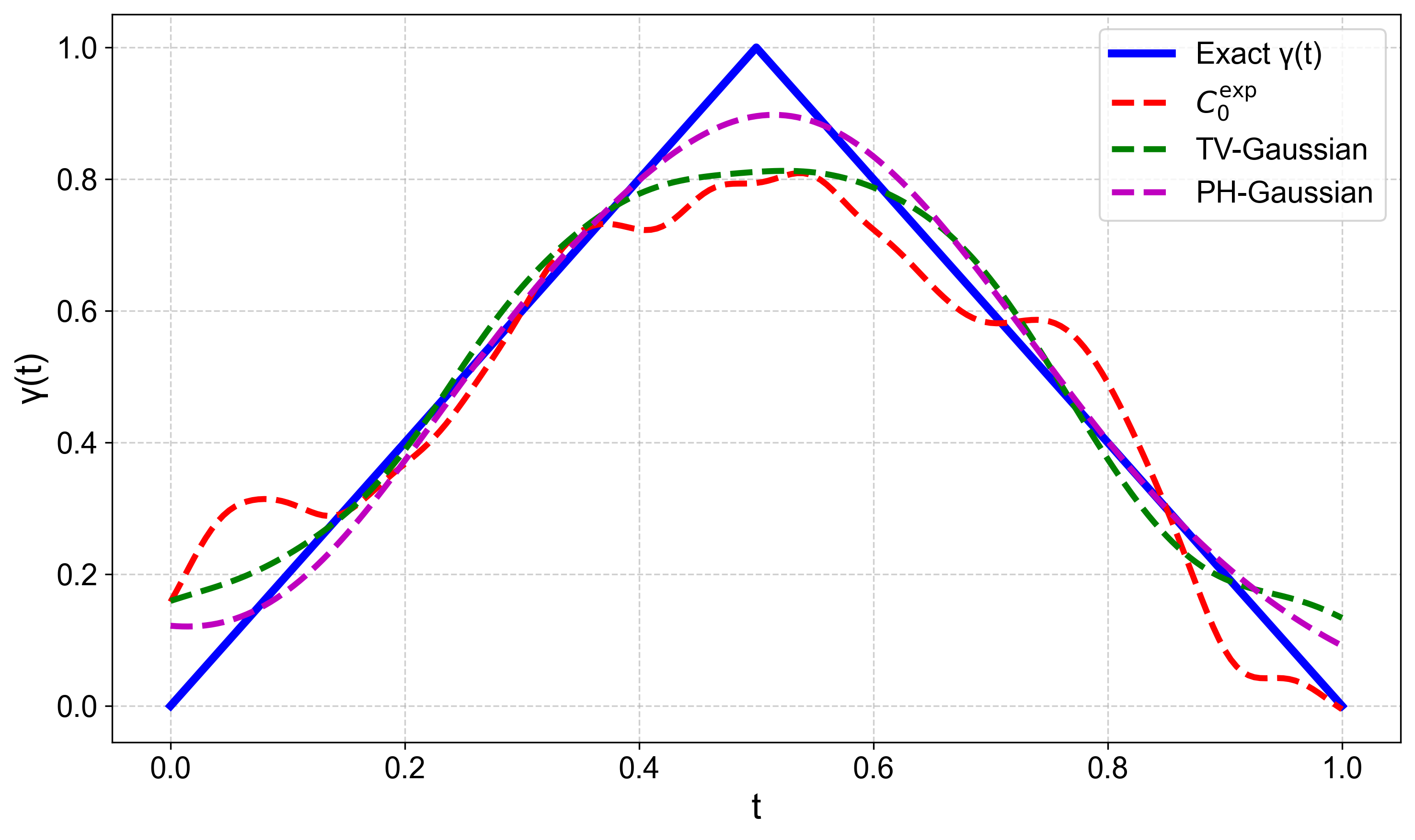}
        \caption{$\epsilon=5\%$}  
        \label{fig:sub2}
    \end{subfigure}
    \caption{The numerical results for Example~\ref{example2} with (a) 1$\%$ and (b) 5$\%$ noise added into the data.}
    \label{fig:main4}  
\end{figure}
\FloatBarrier

\begin{table}[htbp]
    \caption{Numerical results of Example~\ref{example2} with different noise levels}
    \centering
    \begin{tabular}{c|cc|cc|cc}
        \toprule
        $\epsilon$ & \multicolumn{2}{c}{Gaussian Prior} & \multicolumn{2}{c}{TV-Gaussian Prior} & \multicolumn{2}{c}{PH-Gaussian Prior} \\
        \cmidrule(lr){2-3} \cmidrule(lr){4-5} \cmidrule(lr){6-7}
                   & $\hat{\lambda}$ & $e_r$ & $\hat{\lambda}$ & $e_r$ & $\hat{\lambda}$ & $e_r$ \\
        \midrule
        0.01 & / & 0.0857 & 26.34 & 0.0345 & 8.82 & 0.0248 \\
        0.03 & / & 0.1292 & 29.75 & 0.0668 & 9.79 & 0.0529 \\
        0.05 & / & 0.1630 & 33.78 & 0.1163 & 10.29 & 0.0673 \\
        \bottomrule
    \end{tabular}
  \label{table2}
\end{table}
\FloatBarrier

\begin{example}
\label{example3}
The conductivity $\alpha(x)$ is set as $\alpha(x) = e^{x}$, and the exact Robin coefficient $\gamma^{\dagger}(t)$ is given by $\gamma^{\dagger}(t) =  \frac{1}{2}\chi_{\left\{\frac{3}{10} \leq t \leq \frac{7}{10}\right\}}$. The boundary conditions, source term, and initial condition are specified such that the analytical solution to the direct problem \eqref{eqn1} is given by $u(x,t) = (x^{2}+1)\sin(\pi t)$.
\end{example}

In Example \ref{example3}, we examine a discontinuous step function on $[0.3, 0.7]$. This case tests the PH-Gaussian prior's ability to resolve sharp jumps while suppressing Gibbs-like oscillations that typically affect standard regularization.

\begin{figure}[htbp]
    \centering  

    \begin{subfigure}[b]{0.45\textwidth}  
        \centering
        \includegraphics[width=\textwidth]{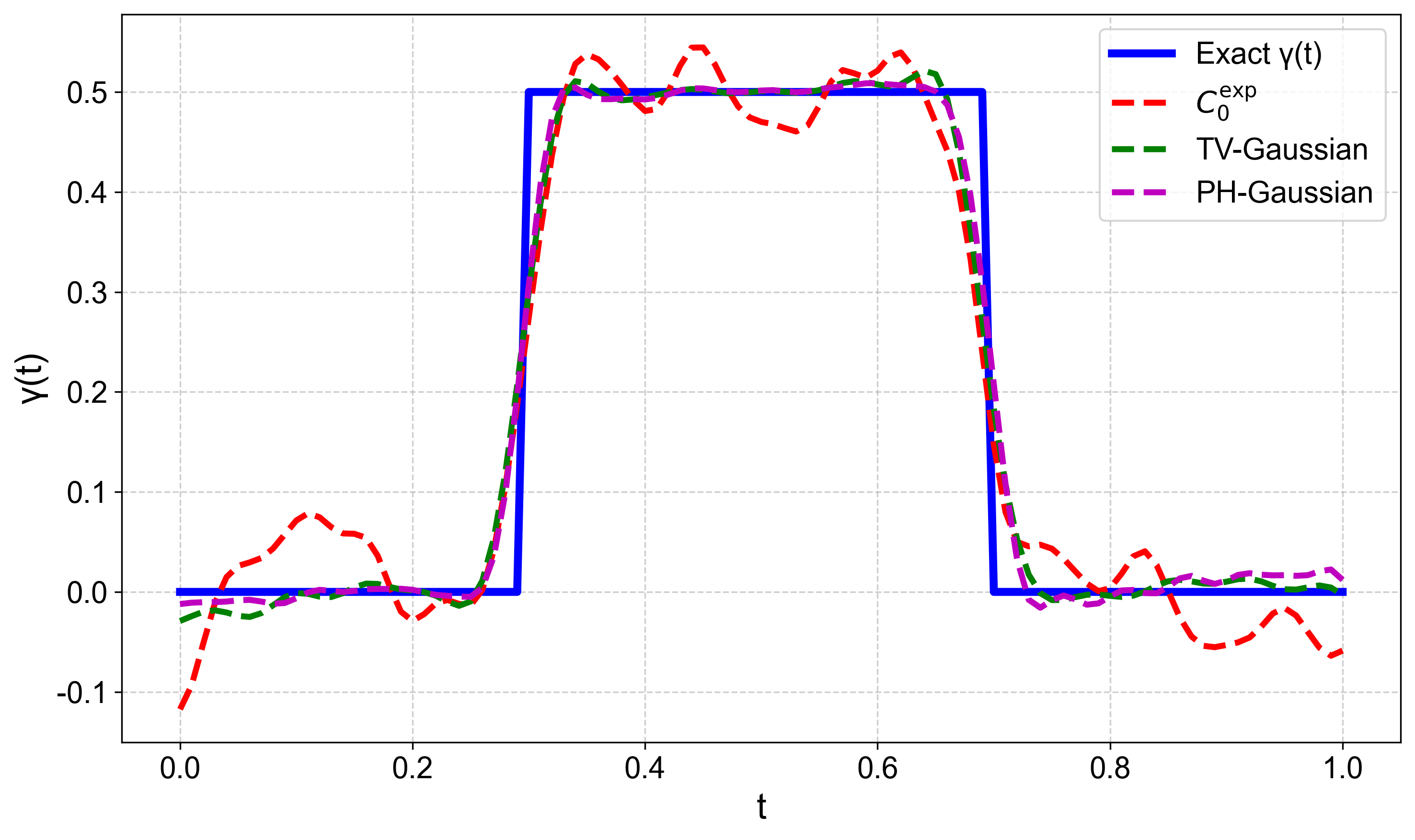}  
        \caption{$\epsilon=1\%$}  
        \label{fig:sub1}  
    \end{subfigure}
    \hspace{0.01\textwidth}  
    \begin{subfigure}[b]{0.45\textwidth}
        \centering
        \includegraphics[width=\textwidth]{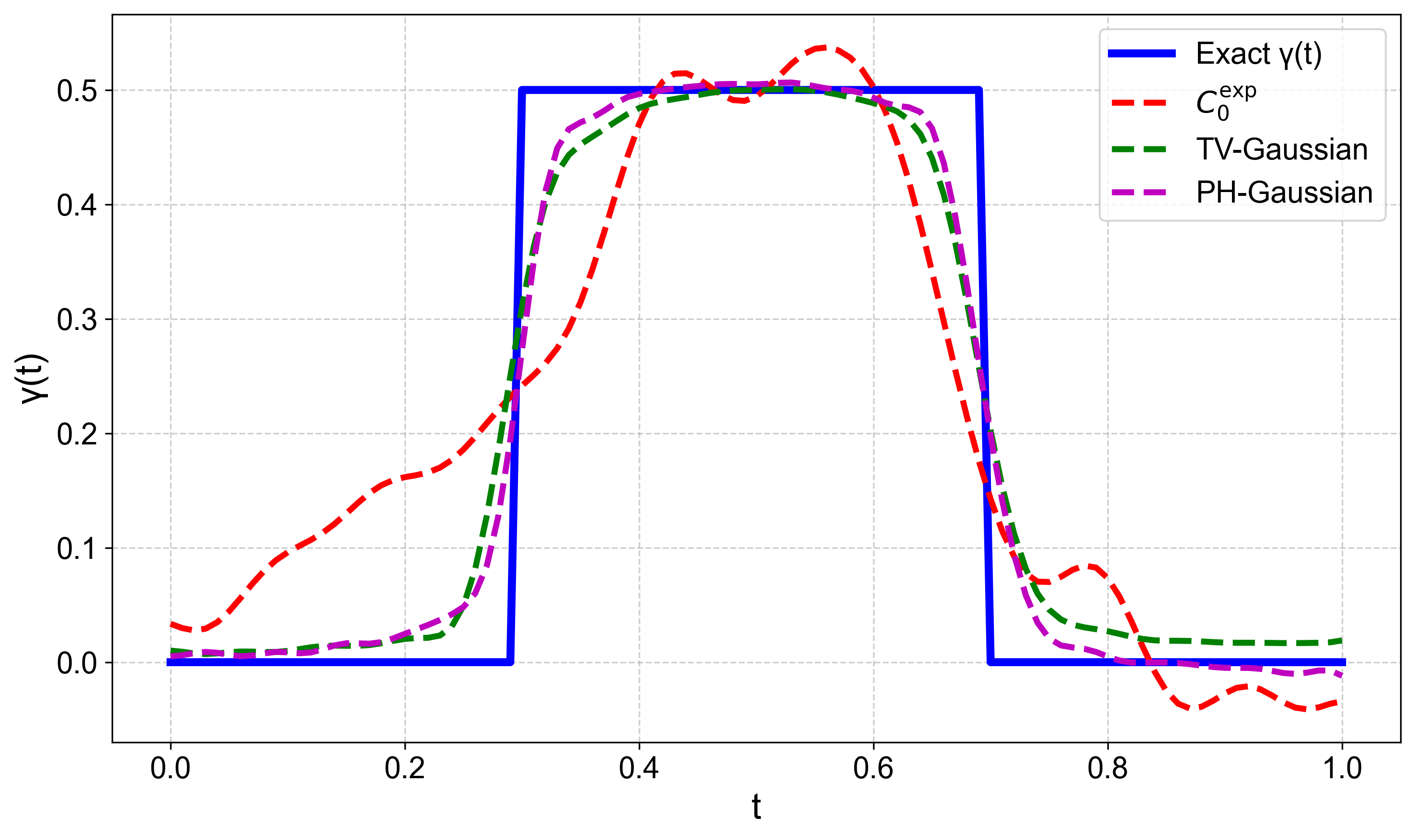}
        \caption{$\epsilon=5\%$}  
        \label{fig:sub2}
    \end{subfigure}

    \caption{The numerical results for Example~\ref{example3}  with (a) 1$\%$ and (b) 5$\%$ noise added into the data.}
    \label{fig:main7}  
\end{figure}

Table \ref{table3} shows that the PH-Gaussian prior maintains the lowest relative errors across all noise levels, outperforming both Gaussian and TV-Gaussian priors. Even at $5\%$ noise, the PH-Gaussian approach ($e_r = 0.1849$) remains more robust than the TV-Gaussian method ($e_r = 0.2086$). Fig. \ref{fig:main7} confirms that the PH-based estimate identifies the step height and location more effectively, avoiding the excessive corner-smoothing seen in the standard Gaussian results.

\begin{figure}[htbp]
    \centering  

    \begin{subfigure}[t]{0.45\textwidth}  
        \centering
        \includegraphics[width=\linewidth]{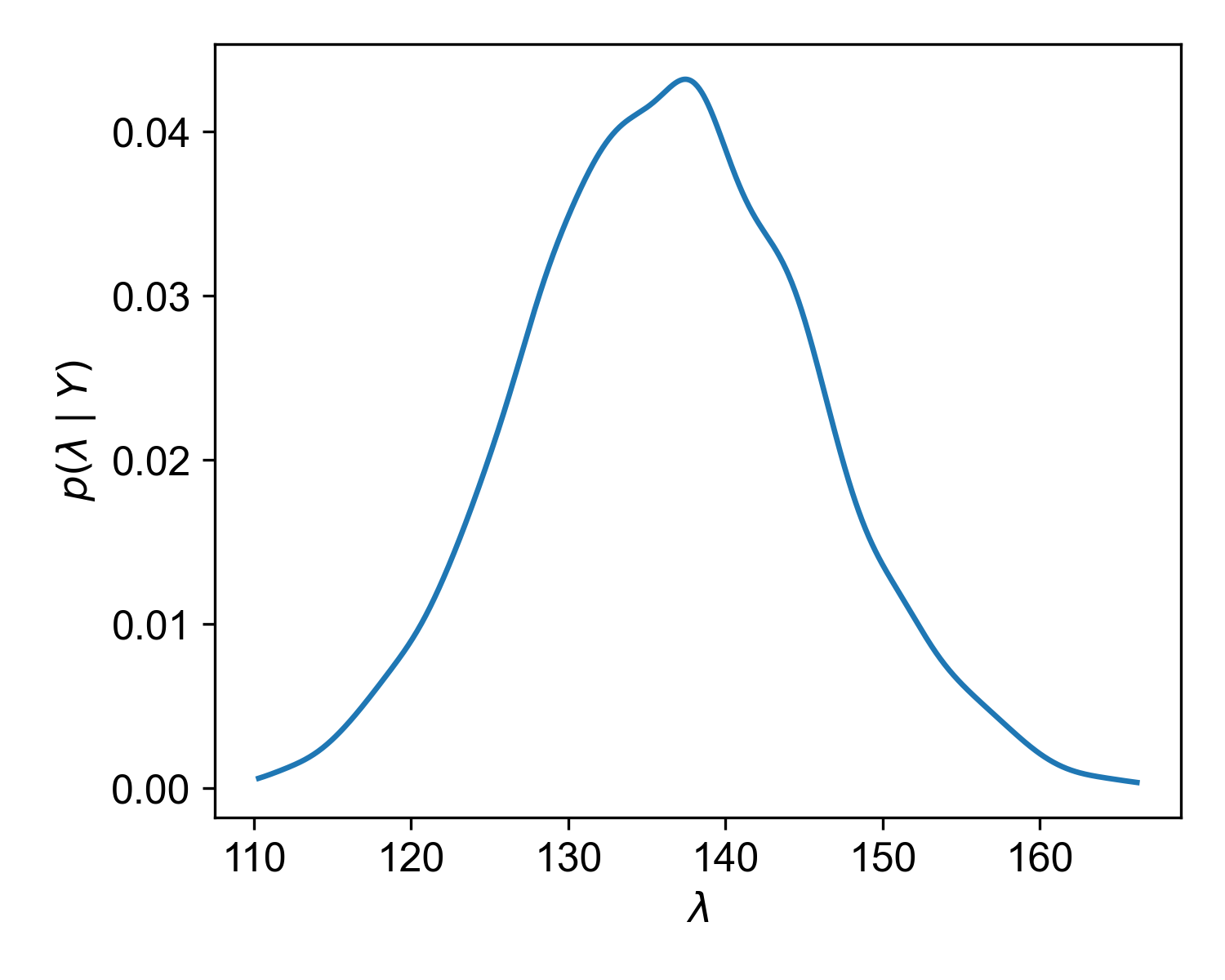}  
        \caption{The posterior density of TV–Gaussian}  
        \label{fig:sub1a}  
    \end{subfigure}
   \hfill
    \begin{subfigure}[t]{0.45\textwidth}
        \centering
        \includegraphics[width=\linewidth]{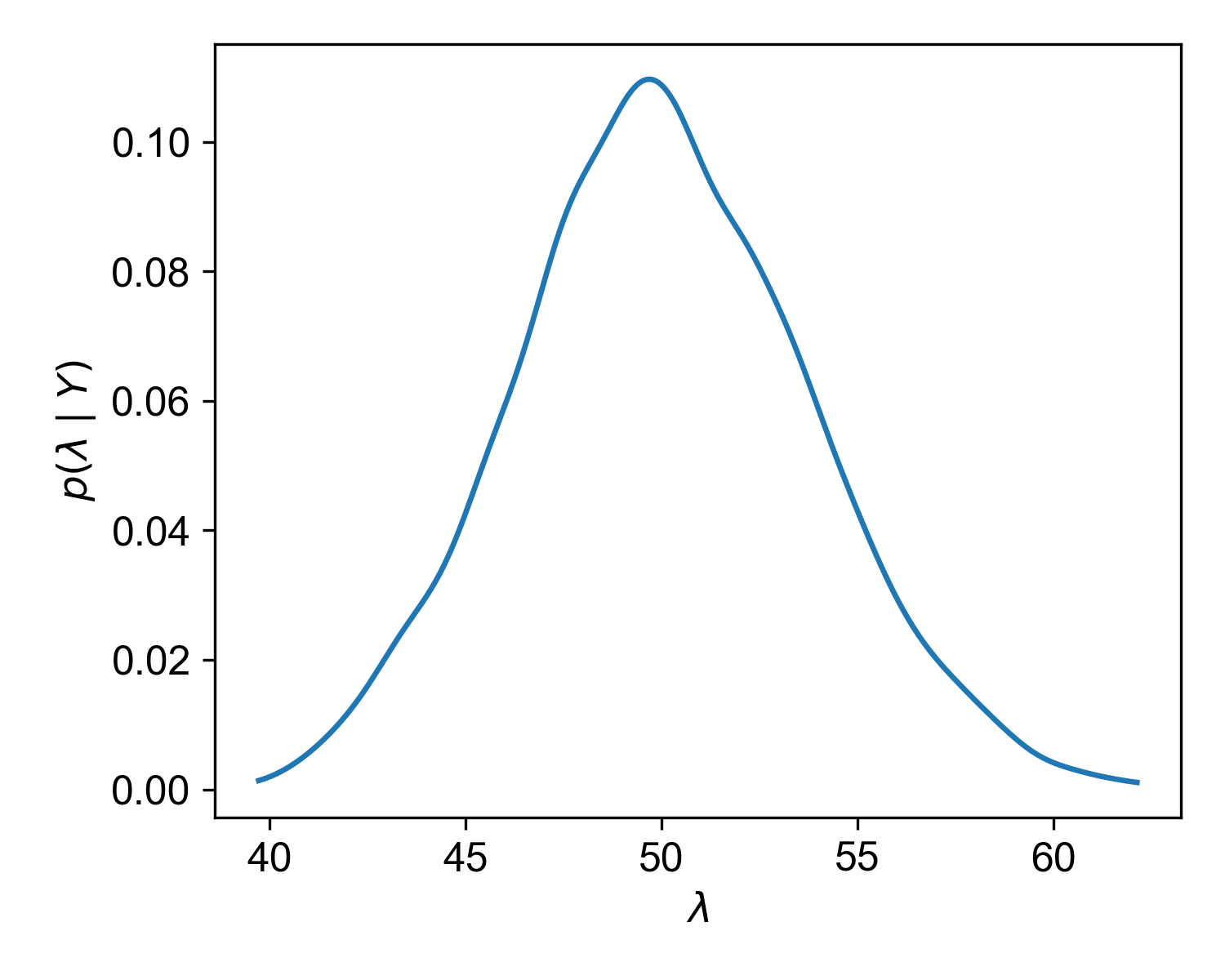} 
        \caption{The posterior density of PH–Gaussian}
        \label{fig:sub2a}
    \end{subfigure}
    \caption{The posterior density of the scaling parameter \(\lambda\) for Example~\ref{example3} with 1$\%$ noise added into the data.}
    \label{fig:main8}  
\end{figure}

The unimodal posterior density of $\lambda$ in Fig. \ref{fig:main8} validates the hierarchical framework's stability for discontinuous targets. The adaptive increase of $\hat{\lambda}$ (from $50.05$ to $58.15$) indicates that the model correctly identifies the need for stronger regularization to preserve the box structure compared to the previous smooth cases.

\begin{figure}[htbp]
    \centering  

    \begin{subfigure}[t]{0.45\textwidth}  
        \centering
        \includegraphics[width=\linewidth]{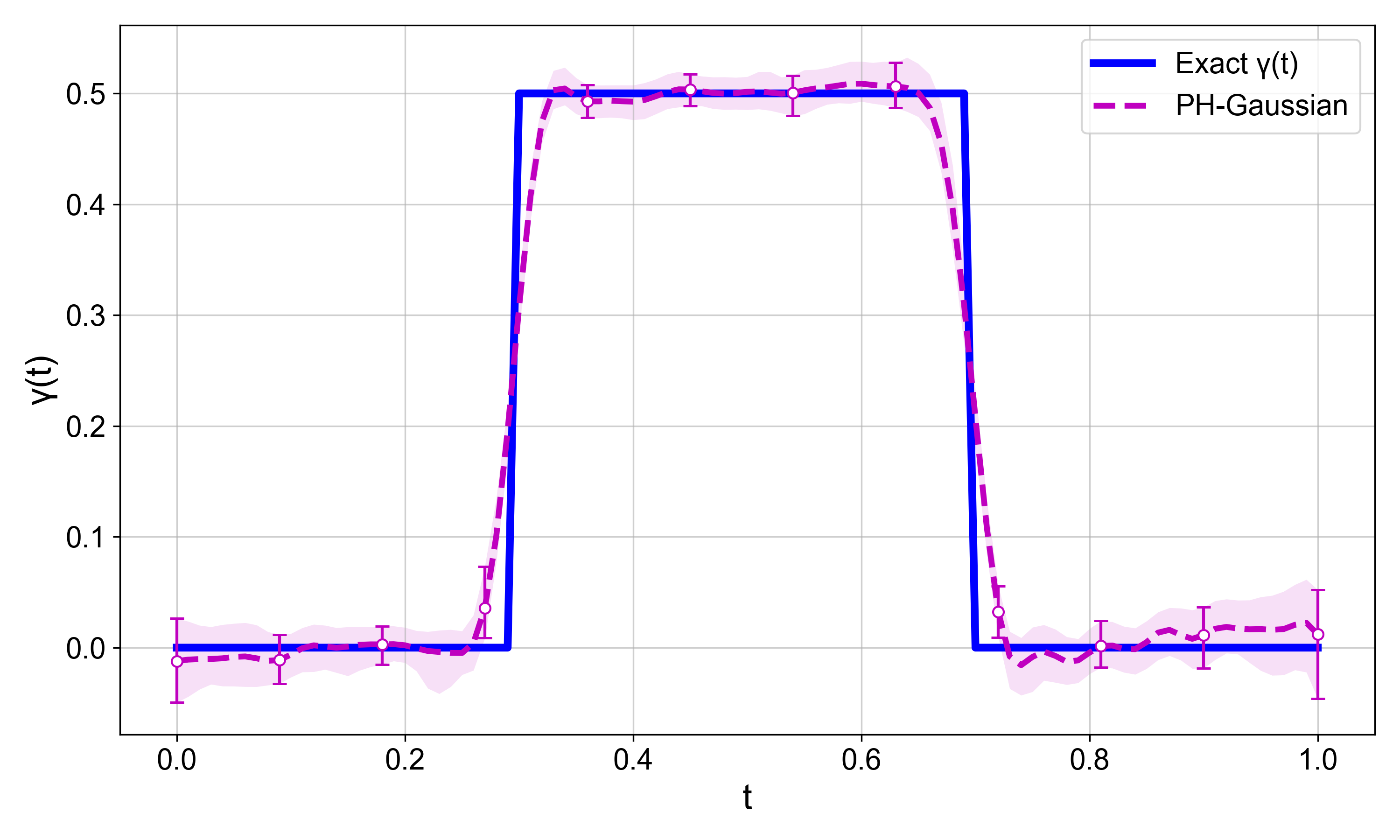}  
        \caption{$\epsilon=1\%$}  
        \label{fig:sub1}  
    \end{subfigure}
 \hfill
    \begin{subfigure}[t]{0.45\textwidth}
        \centering
        \includegraphics[width=\linewidth]{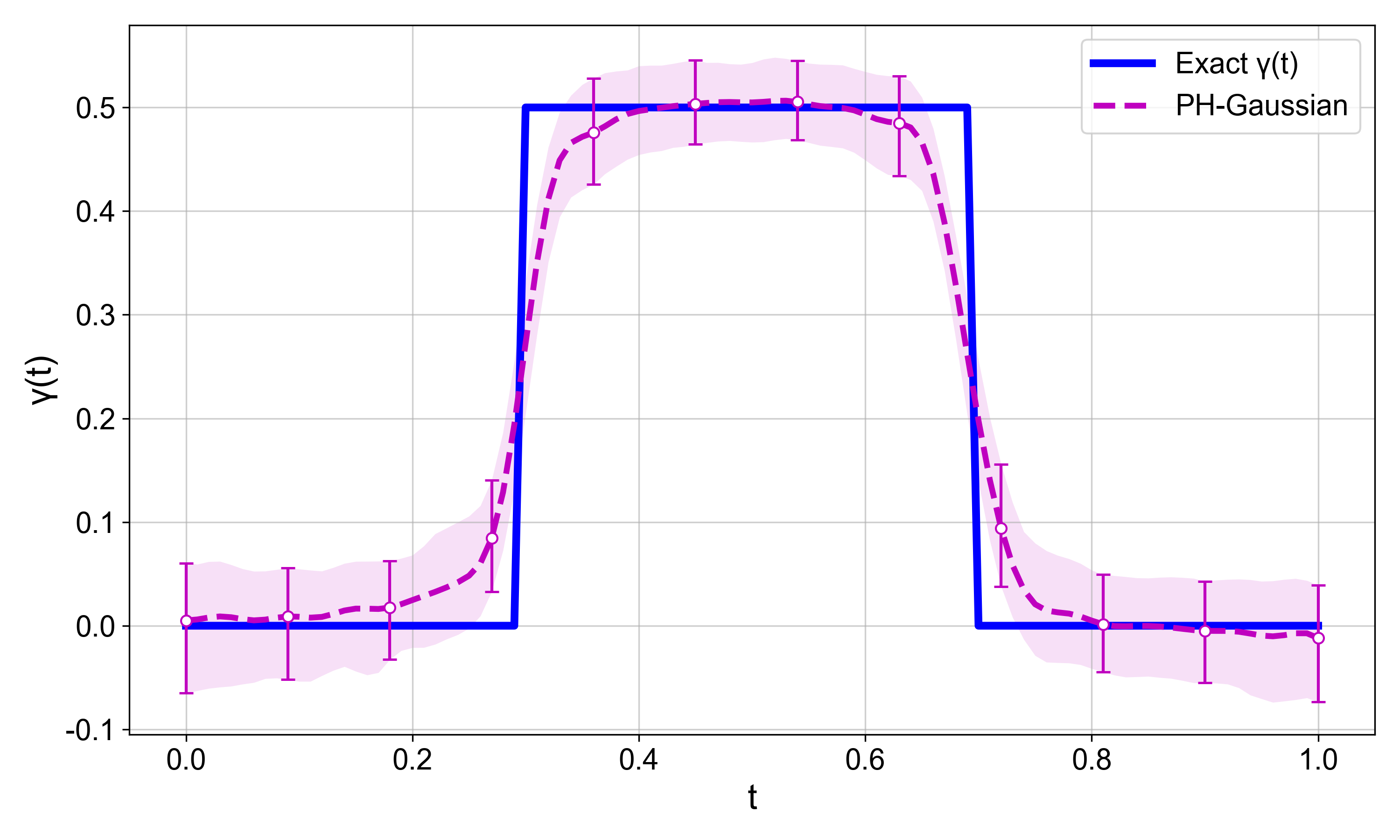}
       \caption{$\epsilon=5\%$}  
        \label{fig:sub2}
    \end{subfigure}

    \caption{The numerical results for Example~\ref{example3} with (a) 1$\%$ and (b) 5$\%$ noise added into the data.}
    \label{fig:main9}  
\end{figure}

Fig. \ref{fig:main9} illustrates the uncertainty quantification for the discontinuous case. With increasing noise, the confidence interval widens significantly near the discontinuities. This objectively reflects the inherent information uncertainty in extracting step edges from noisy data.

\begin{table}[htbp]
    \caption{Numerical results of Example~\ref{example3}  with different noise levels}
    \centering
    \begin{tabular}{c|cc|cc|cc}
        \toprule
        $\epsilon$ & \multicolumn{2}{c}{Gaussian Prior} & \multicolumn{2}{c}{TV-Gaussian Prior} & \multicolumn{2}{c}{PH-Gaussian Prior} \\
        \cmidrule(lr){2-3} \cmidrule(lr){4-5} \cmidrule(lr){6-7}
                   & $\hat{\lambda}$ & $e_r$ & $\hat{\lambda}$ & $e_r$ & $\hat{\lambda}$ & $e_r$ \\
        \midrule
        0.01 & / & 0.1987 & 136.46 & 0.1487 & 50.05 & 0.1432 \\
        0.03 & / & 0.3076 & 162.39 & 0.1874 & 55.33 & 0.1667 \\
        0.05 & / & 0.3611 & 174.06 & 0.2086 & 58.15 & 0.1849 \\
        \bottomrule
    \end{tabular}
\label{table3}
\end{table}

\FloatBarrier

In summary, the PH-Gaussian prior consistently outperforms standard Gaussian and TV-Gaussian priors across smooth, peak, and discontinuous cases. By leveraging persistent homology, the method preserves topological structures and yields lower relative errors under high noise. The hierarchical framework enables automatic, noise-adaptive regularization, while uncertainty quantification provides a statistical measure of reconstruction risks, with credible intervals reflecting localized uncertainty at singular features. These results establish the PH-Gaussian prior as a robust, self-adaptive strategy for thermal inverse problems.

\section{Conclusion}
In this paper, we studied the identification of a time-dependent Robin coefficient in a parabolic heat-transfer problem within a Bayesian framework. A PH–Gaussian prior was adopted to incorporate topological information of the Robin coefficient, and a hierarchical Bayesian method was used to infer the regularization parameter $\lambda$.  Numerical results show that the PH-Gaussian prior performs  well. Also, the hierarchical Bayesian framework can automatically determine suitable regularization parameters. The proposed method offers a novel approach to solving the Robin coefficient identification problem in inverse heat conduction equations.





\bibliographystyle{plain}   
\bibliography{mybib}       
\end{document}